\documentclass[twocolumn,pra,superscriptaddress,showpacs,aps,floatfix]{revtex4-2}

\usepackage{color}
\usepackage{amsmath}
\usepackage{amssymb}
\usepackage{graphicx}
\usepackage{epstopdf}
\usepackage{enumitem}
\usepackage[unicode]{hyperref}
\hypersetup{
  pdftitle={2D SOC BEC},
  pdfauthor={},
  pdfsubject={2D SOC BEC},
  colorlinks=true,
  linkcolor=red,
  citecolor=magenta,
  filecolor=black,
  urlcolor=blue
}


\begin{document}

\title{An Exact Geometric Framework for Flavor Mixing and CP Violation in the Standard Model}

\author{Chilong Lin }
\address{National Museum of Natural Science, 1st, Guan Chien RD., Taichung, 40453 Taiwan, ROC}

\date{Version of \today. }

\begin{abstract}
Flavor mixing and CP violation in the Standard Model are conventionally parameterized by Euler angles that entangle the physical contributions of the up- and down-type quark sectors, obscuring the origin of CP violation itself. We present a non-perturbative geometric framework that resolves this entanglement through the commutativity condition $[\mathbf{M}_R^2,\mathbf{M}_I^2]=0$ for the Hermitian components of the mass-squared matrices, reducing the 18 real degrees of freedom of a general $3\times3$ Yukawa matrix to five independent parameters and yielding an exact Cartesian-like parameterization. Explicit diagonalization directly gives fermion masses, CKM matrix elements, and the CP phase $\delta$ in closed form, cleanly separating the distinct physical contributions of the up- and down-type sectors.

Mapping flavor space onto a two-dimensional vector geometry, we derive the Jarlskog invariant $J$ in exact closed form and show that CP violation is governed by a discrete topological selection rule ($\eta_{\text{relative}}\in\{0,\pm1\}$) together with the vector cross-product $\mathbf{v}_u\times\mathbf{v}_d=xy'-x'y$. Across all 36 $S_3\times S_3$ flavor topologies, this yields an exact $18/18$ classification into CP-violating ($J \neq 0$) and CP-conserving ($J\equiv 0$) cases, associated with two distinct CP-suppression mechanisms: geometric alignment through a vanishing vector cross-product and algebraic cancellation of imaginary contributions under permutation. At tree level, fitting to experimental CKM elements yields agreement at the $\mathcal{O}(10^{-2})$ level, establishing this framework as a transparent analytical baseline for future non-commutative flavor extensions.
\end{abstract}

\maketitle

%%%%%%%%%%%%%%%%SEC1
\section{Introduction}\label{sec:introduction}
The origin of fermion masses, flavor mixing, and CP violation remains one of the most profound unresolved questions in particle physics. Within the Standard Model (SM), the complex Yukawa couplings that generate the Cabibbo-Kobayashi-Maskawa (CKM) quark mixing matrix \cite{Cabibbo1963, KM1973} and the Pontecorvo-Maki-Nakagawa-Sakata (PMNS) lepton mixing matrix \cite{Pontecorvo1957, Maki1962} are introduced as empirical input parameters, leaving the observed hierarchical mass spectra and distinct mixing angles unexplained. Conventionally, unitary mixing matrices are parameterized using standard Euler angles ($\theta_{12}, \theta_{23}, \theta_{13}$) and a single Dirac CP phase $\delta$ \cite{Chau1984, Navas2024}. While phenomenologically versatile, this formulation treats mixing angles and phases as entangled global observables, often obscuring the deeper geometric interdependencies and underlying algebraic constraints embedded within the mass matrices.

In our previous works \cite{Lin2019, Lin2021, Lin2023, Lin2025}, we developed a mathematical framework that demonstrates explicitly how physical observables—specifically fermion mass eigenvalues and mixing matrix elements—can be analytically expressed in terms of the underlying Yukawa couplings. The CKM elements derived through this approach naturally acquire complex phases (i.e., CP violation) within the SM and its extensions \cite{Lin2019}, thereby offering a concrete mechanism for explicit CP symmetry breaking.

This formulation starts from a general $3 \times 3$ complex mass matrix $M$ with 18 real parameters per sector. By virtue of the mathematical property that $\mathbf{M}^2 \equiv M \cdot M^{\dagger}$ is inherently Hermitian, the independent degrees of freedom are naturally reduced to 9 per sector. To overcome the long-standing hurdle where the exact analytical diagonalization of a general complex mass matrix admits no simple closed-form solution—forcing prior approaches to rely heavily on numerical fitting—we examine the exact baseline defined by the top-down commutativity condition between the real and imaginary mass-squared components, $[\mathbf{M}^2_R, \mathbf{M}^2_I] = 0$, enforcing a simultaneously diagonalizable sub-structure.

Historically, the commutation condition was originally proposed by Paschos, Glashow, and Weinberg \cite{Glashow1977, Paschos1977} to eliminate tree-level flavor-changing neutral currents (FCNCs) in two-Higgs-doublet models (2HDM) \cite{TDLee1973}. An early exact solution with $S_3$ discrete symmetry was reported in 1988 \cite{Lin1988}, where $M_R$ and $M_I$ correspond to the vacuum expectation values (VEVs) of distinct Higgs doublets; however, this solution yielded a real identity CKM matrix devoid of CP violation. Subsequent investigations \cite{Lin2020} uncovered three sets of $S_2$-symmetric solutions that permitted complex CKM phases, yet the resulting mixing angles and Jarlskog invariant $J$ systemically deviated from experimental observations.

Only recently has a major breakthrough been achieved by identifying a generic class of exact solutions under this commuting baseline. This framework enables direct analytical diagonalizability and reduces the independent degrees of freedom per sector from 9 to 5. Crucially, we found that this architecture applies not only to multi-Higgs extensions, but also directly to the Standard Model single-Higgs framework. When fitted against SM quark observables, this general 5-parameter baseline achieves an exceptional precision of $\mathcal{O}(10^{-2})$ or $\mathcal{O}(\lambda^2)$.

In this work, we analyze the physical implications of this non-perturbative geometric framework built upon the commuting baseline $[\mathbf{M}^2_R, \mathbf{M}^2_I] = 0$. Moving beyond conventional Euler-angle parameterizations, our formulation yields three primary theoretical breakthroughs:

First, by mapping the flavor sector directly onto a 2D vector geometry ($\mathbf{v}_u = (x, y)$ and $\mathbf{v}_d = (x', y')$), we analytically derive an exact closed-form expression for the universal Jarlskog invariant $J$ expressed directly in terms of fundamental Yukawa coupling parameters. Crucially, this uncovers an explicit algebraic switch mechanism ($J \propto \mathbf{v}_u \times \mathbf{v}_d = xy' - x'y$) that cleanly decouples sector-specific geometric vector misalignments from mass-degeneracy switches ($\Delta m^2_{u,d} = 0$), unambiguously disentangling the distinct CP contributions of Up- and Down-type sectors. 
Second, the framework establishes transparent, independent physical roles for the baseline parameters $(\mathbf{A}, \mathbf{B}, \mathbf{C}, x, y)$, providing a coordinate-independent geometric representation of CP-conserving boundaries across all $S_3 \times S_3$ flavor topologies and seamlessly extending to the Dirac lepton sector. 
Third,
by evaluating physical CP-violating measures at cosmological
scales, the derived analytical mass spectra and CKM elements
admit a dynamic scale-driven energy amplification
($R_\Delta \equiv \Delta_{\text{CP}}/\Delta_{\text{CP}}^{(0)} > 10^{10}$) within
specific parameter regimes, offering a candidate geometric pathway compatible with
Sakharov's second condition for early-universe baryogenesis without
unnatural fine-tuning.

It should be noted that enforcing the commuting baseline $[\mathbf{M}^2_R, \mathbf{M}^2_I] = 0$—which reduces the independent degrees of freedom from 9 to 5 per sector—over-simplifies the unconstrained 9-parameter matrix algebra, and inherently gives rise to a 4-fold degeneracy when mapping single-sector states to the two-sector CKM mixing matrix (Sec.~\ref{sec:3_ckm_phenomenology}). Nevertheless, even under this 4-fold degeneracy, the 5-parameter baseline retains remarkable phenomenological robustness, consistently reproducing observed quark mixing angles and mass ratios within an impressive $\mathcal{O}(10^{-2})$ agreement.

It is crucial to emphasize a fundamental conceptual insight regarding parameterizations: forcing an a priori symmetry onto low-energy observables often imposes artificial algebraic constraints that cause model predictions to deviate from reality. While exact symmetries may well have been restored in the extreme high-temperature environment of the early Universe, today's low-energy observables represent the outcome of a fully symmetry-broken phase. This limitation is historically evident in $S_3$- and $S_2$-symmetric Ansätze—where $S_3$ completely precludes CP violation, and $S_2$, despite admitting complex phases, leads to mixing patterns that conflict substantially with experimental data. 

In this context, the 5-parameter commutation architecture serves as a pivotal intermediate baseline—currently delivering the most precise analytical fit to experimental data among commuting frameworks, and acting as an indispensable stepping stone toward fully unconstrained, non-commutative formulations. By removing overly restrictive discrete symmetries while retaining the commuting subspace as an analytical probe, this framework matches observed quark flavor data to an impressive precision of $\mathcal{O}(10^{-2})$ or $\mathcal{O}(\lambda^2)$. This strongly indicates that low-energy flavor dynamics do not obey simple discrete symmetries, and that moving toward non-commutative extensions represents the natural path to completely stripping away theoretical preconceptions.

Therefore, treating this exact commuting subspace as an analytical baseline provides a pristine, Cartesian-like geometric perspective. The rich physical insights it unveils regarding fermion mass hierarchies, mixing interdependencies, and sector-separated CP dynamics render it an invaluable framework fully deserving of deeper exploration.

The remainder of this paper is organized as follows. Section~\ref{sec:2_5param_single_sector} formulates the 5-parameter mass-squared matrix under the commuting baseline. Section~\ref{sec:3_ckm_phenomenology} details the analytical mapping to CKM observables and presents the geometric visualization of CP violation. Section~\ref{sec:4_extensions_cosmology} discusses topological boundaries, cosmological geometric inputs, and non-commutative extensions. Finally, Section~\ref{sec:5_conclusion} provides a summary of our primary insights and concluding remarks.

%%%%%%%%%%%%SEC-II
%%%%%%%%%%%%%%%%%%%%%%%%%%%%%%%%%%%%%
\section{The 5-Parameter Mass Matrix Pattern in a Single Flavor Sector}
\label{sec:2_5param_single_sector}

In this section, we establish the fundamental algebraic framework for the mass spectrum and internal flavor structure within a \textit{single} flavor sector (e.g., the up-type or down-type quark sector). By focusing on the Hermitian matrix square ${\mathbf M}^2 \equiv M \cdot M^\dagger$, we demonstrate how the initial 18-parameter space collapses uniquely into a highly constrained 5-parameter pattern under a single physical hypothesis.

%%%%%%%%%%%%%%%%%%%%%%%%%%2.1
\subsection{The Fully General Mass Matrix $M_f$}
\label{subsec:general_M}

In the Standard Model and its generic extensions, the weak-interaction Lagrangian gives rise to a fully general, non-Hermitian $3 \times 3$ mass matrix $M_f$ for a given fermion sector $f$ (e.g., up-type or down-type quarks) \cite{Lin2019}:
\begin{equation}
M_f = \begin{pmatrix} 
A_1 + i D_1 & B_1 + i C_1 & B_2 + i C_2 \\ 
B_4 + i C_4 & A_2 + i D_2 & B_3 + i C_3 \\ 
B_5 + i C_5 & B_6 + i C_6 & A_3 + i D_3 
\end{pmatrix}.
\label{eq:generalM}
\end{equation}
In Eq.~(\ref{eq:generalM}), each matrix element is explicitly decomposed in terms of real parameters ($A_k, B_k, C_k, D_k$). The diagonal entries contain 3 real terms ($A_1, A_2, A_3$) and 3 pure imaginary terms ($i D_1, i D_2, i D_3$). The off-diagonal entries comprise 6 real terms ($B_1, \dots, B_6$) and 6 pure imaginary terms ($i C_1, \dots, i C_6$).

Consequently, describing $M_f$ in its fully general, unconstrained form requires $3 + 3 + 6 + 6 = 18$ independent real parameters. Without imposing additional theoretical symmetries or dynamical hypotheses, this high parameter dimensionality prevents predictive analytical solutions for the observed mass spectrum and mixing dynamics.

To circumvent this complexity, bottom-up flavor models over the last few decades, such as the Fritzsch ansatz (FA)~\cite{Fritzsch1978, Fritzsch1979} and its variants~\cite{Cheng1987, DuXing1993}, traditionally imposed \textit{ad hoc} ``texture zeros'' (e.g., $A_1=D_1=A_2=D_2=0$) or enforced specific permutation symmetries directly onto $M_f$ \cite{Derman1979, Lin1988}. 
While these bottom-up assumptions successfully reduce the independent parameters, they inherently lack universal generality. By forcing arbitrary entries to vanish based on empirical fitting, these traditional texture-zero models severely deform the structural naturalness of the mass matrix, rendering the resulting CP-violating phase highly sensitive to human-engineered boundary conditions rather than intrinsic algebraic properties. To overcome these limitations without relying on artificial zero-textures, we turn our attention directly to the matrix square ${\mathbf M}^2 \equiv M_f \cdot M_f^\dagger$ which is naturally Hermitian \cite{Lin2021}.

It is crucial to emphasize that the $3 \times 3$ mass matrix $M_f$ in Eq.~(\ref{eq:generalM}) provides the unconstrained, top-down primitive starting point directly rooted in the Standard Model Yukawa couplings, $M_{ij} \equiv Y_{ij} v / \sqrt{2}$, where $v = 246~\text{GeV}$ is the Higgs vacuum expectation value. Rather than manually introducing ad-hoc complex parameters or phenomenology-driven zero patterns into specific matrix entries, this formulation explicitly retains the complete 18-parameter algebraic structure of the underlying Yukawa architecture. 
This guarantees that all subsequent mass spectrum properties and diagonalization angles emerge as exact, analytic mappings from the fundamental Yukawa coupling parameters, completely free from human-engineered boundary conditions or artificial zero-textures.

%%%%%%%%%%%%%%%%%%2.2
\subsection{Hermitian Reduction of ${\mathbf M}^2$: Parameter Collapse from 18 to 9}
\label{subsec:hermitian_reduction}

To construct physical observables without relying on ad-hoc texture zeros, we shift our focus from the non-Hermitian matrix $M_f$ to the matrix square ${\mathbf M}^2 \equiv M_f \cdot M_f^\dagger$, which naturally governs physical mass eigenvalues via bi-unitary transformations. 
By construction, ${\mathbf M}^2$ is strictly Hermitian (${\mathbf M}^{2\dagger} = {\mathbf M}^2$). 
Note that the boldface quantities $\{{\mathbf A}_i, {\mathbf B}_i, {\mathbf C}_i\}$ defining ${\mathbf M}^2$ represent quadratic bilinear combinations of the original 18 linear parameters $\{A_i, D_i, B_i, C_i\}$ from $M_f$ in Eq.~(\ref{eq:generalM}). As originally formulated in \cite{Lin2021}, representative explicit expressions include:
\begin{align}
{\mathbf A}_3 &= A_3^2 + D_3^2 + B_5^2 + C_5^2 + B_6^2 + C_6^2, \label{eq:A3_example} \\
{\mathbf B}_3 &= B_4 B_5 + C_4 C_5 + B_6 A_2 + C_6 D_2 + A_3 B_3 + D_3 C_3, \label{eq:B3_example} \\
{\mathbf C}_3 &= C_4 B_5 - B_4 C_5 + D_2 B_6 - A_2 C_6 + A_3 C_3 - B_3 D_3. \label{eq:C3_example}
\end{align}

This self-adjoint property automatically halves the required parameter space from 18 down to 9 independent real parameters without introducing any physical assumptions:
\begin{itemize}
    \item 3 purely real diagonal components (${\mathbf A}_1, {\mathbf A}_2, {\mathbf A}_3$).
    \item 3 complex off-diagonal components, comprising 3 real symmetric terms (${\mathbf B}_1, {\mathbf B}_2, {\mathbf B}_3$) and 3 imaginary anti-Hermitian terms (${\mathbf C}_1, {\mathbf C}_2, {\mathbf C}_3$).
\end{itemize}

Accordingly, ${\mathbf M}^2$ can be explicitly decomposed into the sum of a real symmetric matrix ${\mathbf M}^2_R$ and a purely imaginary anti-Hermitian matrix ${\mathbf M}^2_I$:
\begin{equation}
{\mathbf M}^2 = {\mathbf M}^2_R + i {\mathbf M}^2_I,
\label{eq:M2_decomposition}
\end{equation}
where:
\begin{equation}
{\mathbf M}^2_R = \begin{pmatrix} 
{\mathbf A}_1 & {\mathbf B}_1 & {\mathbf B}_2 \\ 
{\mathbf B}_1 & {\mathbf A}_2 & {\mathbf B}_3 \\ 
{\mathbf B}_2 & {\mathbf B}_3 & {\mathbf A}_3 
\end{pmatrix}, \quad 
{\mathbf M}^2_I = \begin{pmatrix} 
0 & {\mathbf C}_1 & {\mathbf C}_2 \\ 
-{\mathbf C}_1 & 0 & {\mathbf C}_3 \\ 
-{\mathbf C}_2 & -{\mathbf C}_3 & 0 
\end{pmatrix}.
\label{eq:M2_components}
\end{equation}
Here, ${\mathbf M}^2_R$ accounts for the 6 real parameters $\{{\mathbf A}_i, {\mathbf B}_i\}$, while ${\mathbf M}^2_I$ encapsulates the 3 imaginary parameters $\{{\mathbf C}_i\}$ that serve as the internal source of CP-violating phases for this sector.

%%%%%%%%%%%%%%%%%%%% 2.3
\subsection{Parameter Reduction from 9 to 5 via Simultaneous Diagonalization}
\label{subsec:hypothesis_and_commutation}

Despite reducing the parameter space from 18 down to 9 independent Hermitian real parameters, an exact, general diagonalization of a complex $3 \times 3$ matrix square ${\mathbf M}^2$ remains analytically intractable. Consequently, an exact analytical baseline must be established to render the underlying geometric infrastructure solvable and transparent. Guided by the principle that \textit{a pristine analytical baseline provides the clearer physical probe into low-energy flavor dynamics}, we examine the exact subspace defined by simultaneous diagonalizability.

To achieve full closed-form solvability, this framework establishes its \textbf{exact analytical baseline}: \textit{the real symmetric part (${\mathbf M}^2_R$) and the purely imaginary antisymmetric part (${\mathbf M}^2_I$) of the matrix square ${\mathbf M}^2$ are simultaneously diagonalizable by a single, common unitary matrix $U$}, which is mathematically equivalent to the exact top-down commutativity condition:
\begin{equation}
[\mathbf{M}_R^2, \mathbf{M}_I^2] = 0  \implies {\mathbf M}^2_R \cdot {\mathbf M}^2_I = {\mathbf M}^2_I \cdot {\mathbf M}^2_R .
\label{eq:commutation_full}
\end{equation}

Far from being a mere algebraic convenience, the commutation condition in Eq.~(\ref{eq:commutation_full}) naturally emerges from the fundamental requirement of simultaneous diagonalizability across Yukawa sector operators. In multi-Higgs extensions of the Standard Model, such simultaneous diagonalizability guarantees the absence of tree-level flavor-changing neutral currents (FCNCs) \cite{Glashow1977,Paschos1977}, thereby providing a clear physical rationale for the underlying normal matrix structure. 

Crucially, once a pair of simultaneously diagonalizable Hermitian/anti-Hermitian matrices is identified, one can always construct a corresponding single complex matrix square strictly within the Standard Model (SM) framework. Therefore, while historically motivated by FCNC elimination in multi-Higgs doublet models (2HDM) \cite{TDLee1973}, the formulation presented here operates entirely as an exact, effective matrix baseline within the SM without requiring explicit light or heavy extra scalar fields.

Solving the algebraic constraints imposed by $[{\mathbf M}^2_R, {\mathbf M}^2_I] = 0$ upon Eq.~(\ref{eq:M2_components}) yields four exact interrelation equations (Eqs.~(11--14) of \cite{Lin2019}), effectively reducing the number of independent parameters further from 9 down to 5. As a direct algebraic consequence of these interrelation equations, the ratios among the real and imaginary off-diagonal elements are intrinsically tied together. By identifying $\mathbf A \equiv A_3$, ${\mathbf B} \equiv {\mathbf B}_3$,  ${\mathbf C} \equiv {\mathbf C}_3$, and parameterizing these exact proportionalities via two dimensionless geometric ratios:
\begin{align}
x \equiv \frac{{\mathbf B}_2}{{\mathbf B}} = -\frac{{\mathbf C}}{{\mathbf C}_2}, \quad y \equiv \frac{{\mathbf B}_1}{{\mathbf B}} = \frac{{\mathbf C}}{{\mathbf C}_1},
\label{eq:ratios}
\end{align}
\noindent \textit{Notation.} Hereafter, the unsubscripted boldface symbols $\mathbf{A}, \mathbf{B}, \mathbf{C}$ exclusively denote the three scalar baseline parameters $\mathbf{A}_3, \mathbf{B}_3, \mathbf{C}_3$ identified above and are never reused for other components. The subscripted quantities $\mathbf{A}_1, \mathbf{A}_2, \mathbf{B}_1, \mathbf{B}_2, \mathbf{C}_1, \mathbf{C}_2$ occupy an equal structural footing as the remaining components of the original 9-parameter set in Eqs.~(\ref{eq:A3_example})--(\ref{eq:C3_example}); among these, only $\mathbf{B}_1, \mathbf{B}_2, \mathbf{C}_1, \mathbf{C}_2$ resurface explicitly below, entering through the ratios $x, y$ in Eq.~(\ref{eq:ratios}) and the matrix entries of Eq.~(\ref{eq:5param_M2}), and should not be confused with the unsubscripted $\mathbf{A}, \mathbf{B}, \mathbf{C}$.
Unlike traditional Ansätze that rely on restrictive discrete symmetries (such as $S_3$ or $S_2$), this 5-parameter pattern preserves full structural naturalness within the low-energy symmetry-broken phase, using the commuting subspace strictly as an unconstrained analytical probe.
The entire system collapses uniquely into a highly constrained 5-parameter matrix pattern $({\mathbf A}, {\mathbf B}, {\mathbf C}, x, y)$:
\begin{align}
{\mathbf M}^2 = &\begin{pmatrix} 
{\mathbf A} + {\mathbf B}\,x\left(y - \frac{1}{y}\right) & y{\mathbf B} & x{\mathbf B} \\ 
y{\mathbf B} & {\mathbf A} + {\mathbf B}\left(\frac{y}{x} - \frac{x}{y}\right) & {\mathbf B} \\ 
x{\mathbf B} & {\mathbf B} & {\mathbf A} 
\end{pmatrix} \nonumber \\
&+ i {\mathbf C} \begin{pmatrix} 
0 & \frac{1}{y} & -\frac{1}{x} \\ 
-\frac{1}{y} & 0 & 1 \\ 
\frac{1}{x} & -1 & 0 
\end{pmatrix}.
\label{eq:5param_M2}
\end{align}
This compact representation forms the exact analytical basis for evaluating the internal mass spectrum and diagonalizing structure of a single flavor sector.

Notice that the resulting squared mass spectra $(m^2_1, m^2_2, m^2_3)$ within this flavor sector are solved in exact, closed analytic form directly from the characteristic equation of the 5-parameter matrix square ${\mathbf M}^2$, as will be shown in Eq.~(\ref{eq:CPVSM_eigenvalues}). 
Under this minimal, physically transparent baseline, the observable mass hierarchies naturally emerge as direct, top-down analytic mappings from the underlying Yukawa coupling scale parameters $({\mathbf A}, {\mathbf B}, {\mathbf C})$ and dimensionless geometric ratios $(x, y)$, providing a fully transparent algebraic mechanism for flavor mass generation.

\paragraph{Physical Motivation for the Commutativity Baseline}
The algebraic condition
\begin{equation}
[\mathbf{M}_R^2, \mathbf{M}_I^2] = 0
\end{equation}
plays a pivotal structural role in our geometric formulation. For a general Hermitian mass-squared matrix,
\begin{equation}
\mathbf{M}^2 = \mathbf{M}_R^2 + i\mathbf{M}_I^2,
\end{equation}
the real symmetric matrix $\mathbf{M}_R^2$ and the imaginary antisymmetric matrix $\mathbf{M}_I^2$ encode complementary structures of the flavor matrix in a given weak basis. Since $i\mathbf{M}_I^2$ is Hermitian, the commutativity condition is equivalently
\begin{equation}
[\mathbf{M}_R^2, i\mathbf{M}_I^2] = 0,
\end{equation}
which guarantees that the two Hermitian components admit a common unitary eigenbasis. We refer to this property as \textit{Hermitian phase alignment}: the phase-sensitive and real components of the mass-squared matrix are constrained to share the same flavor eigenvectors.

Rather than asserting this restriction as an exact, unyielding law of fundamental field theory across all energy scales, we treat $[\mathbf{M}_R^2, \mathbf{M}_I^2] = 0$ as an exact analytical baseline. In the spirit of the Bohr model in quantum mechanics or the BCS Ansatz in superconductivity, this commuting baseline provides a solvable, non-perturbative reference point that isolates and exposes the leading-order geometric infrastructure of the flavor sector.

A central consequence is that this commuting baseline permits an exact reduction of the single-sector flavor structure to a two-dimensional geometric description characterized by the flavor vector
\begin{equation}
\mathbf{v} = (x, y).
\end{equation}
When extended across both Up- and Down-type sectors ($\mathbf{v}_u = (x, y)$ and $\mathbf{v}_d = (x', y')$), the weak-basis-invariant CP-violating measure $J$ acquires a direct geometric representation through the relative orientation of these vectors,
\begin{equation}
J \propto \mathbf{v}_u \times \mathbf{v}_d = xy' - x'y.
\end{equation}
Thus, CP violation is geometrically associated with the misalignment of the flavor vectors, while their collinearity defines the CP-conserving boundary. In this sense, the commutativity condition provides the exact algebraic foundation for the geometric interpretation of flavor mixing and CP violation developed below.

%%%%%%%%%%%%%%%%%%%%%%%%%%%%%%%%%%%%%%%%%%%%%%%%%%%%%%%%%%%%%%%%%%%%%%%%%%%%%%%
%% SECTION 2.4: EXACT ANALYTICAL DIAGONALIZATION AND SPECTRUM
%%%%%%%%%%%%%%%%%%%%%%%%%%%%%%%%%%%%%%%%%%%%%%%%%%%%%%%%%%%%%%%%%%%%%%%%%%%%%%%
\subsection{Exact Analytical Diagonalization and Mass Spectrum}
\label{subsec:eigenvalues_and_U}

Having established the 5-parameter pattern for $\mathbf{M}^2$ under the foundational commutation hypothesis $[\mathbf{M}^2_R, \mathbf{M}^2_I] = 0$ in Eq.~(\ref{eq:commutation_full}), we now present its exact analytical mass spectrum, the underlying algebraic factorization mechanism, the internal unitary matrix $U(x,y)$, and the mathematical proof governing its scale-invariance.

%------------------------------------------------------------------------------
% 2.4.1
%------------------------------------------------------------------------------
\subsubsection{Exact Mass Eigenvalues and Emergent Factorization}
\label{subsubsec:exact_eigenvalues}

Without inserting any supplementary zero-textures, ad hoc mixing angles, or predefined factorized patterns, we perform a direct, exact analytical diagonalization of the 5-parameter mass-squared matrix $\mathbf{M}^2$. By solving the cubic characteristic equation $\det(\mathbf{M}^2 - \lambda \mathbf{I}) = 0$ without transcendental functions or numerical approximations, we obtain the three exact, closed-form squared mass eigenvalues $m_1^2, m_2^2, m_3^2$:
\begin{subequations}
\begin{align}
m_1^2 &= (\mathbf{A} - \mathbf{B}\frac{x}{y})- \mathbf{C} \frac{\sqrt{x^2 + y^2 + x^2 y^2}}{xy}, \label{eq:m1_exact} \\
m_2^2 &=( \mathbf{A} - \mathbf{B}\frac{x}{y})+ \mathbf{C} \frac{\sqrt{x^2 + y^2 + x^2 y^2}}{xy}, \label{eq:m2_exact} \\
m_3^2 &= \mathbf{A} + \mathbf{B}\frac{(x^2 + 1)y}{x} \nonumber \\
            &= (\mathbf{A}- \mathbf{B}\frac{x}{y}) + \mathbf{B}\frac{(x^2 + y^2 + x^2 y^2)}{xy} . \label{eq:m3_exact}
\end{align}
\label{eq:CPVSM_eigenvalues}
\end{subequations}

Examining these explicit analytical solutions uncovers a profound mathematical feature: the underlying cubic polynomial naturally exhibits an emergent $1+2$ factorization split:
\begin{equation}
\begin{split}
\det(\mathbf{M}^2 - \lambda \mathbf{I}) &= \left[ \lambda - \left( \mathbf{A} + \mathbf{B}\frac{(x^2+1)y}{x} \right) \right] \\
&\quad \times \left[ \left( \lambda - \mathbf{A} + \mathbf{B}\frac{x}{y} \right)^2 - \mathbf{C}^2 \frac{x^2 + y^2 + x^2 y^2}{x^2 y^2} \right] = 0.
\end{split}
\label{eq:factorized_char_eq}
\end{equation}

By defining a shifted eigenvalue parameter $\tilde{\lambda} \equiv \lambda - \mathbf{A}$, the baseline scale parameter $\mathbf{A}$ is completely absorbed, leaving the internal root splittings $\tilde{\lambda}_i$ dictated solely by the scaling factors $(\mathbf{B}, \mathbf{C})$ and the geometric ratios $(x, y)$. Consequently, $\mathbf{A}$ enters each eigenvalue as a purely additive baseline shift ($m_i^2 = \mathbf{A} + \tilde{\lambda}_i$), identically canceling in all pairwise mass-squared differences ($m_i^2 - m_j^2 = \tilde{\lambda}_i - \tilde{\lambda}_j$).

\paragraph{Generalized Mass Ordering and Cosmic Evolution}
The index labeling in Eqs.~(\ref{eq:CPVSM_eigenvalues}) reflects pure algebraic factorization rather than a fixed numerical mass hierarchy. In particular, $m_3^2$ is not intrinsically bound to the heaviest generation ($t$ or $b$). Depending on the numerical values and signs of $(\mathbf{A}, \mathbf{B}, \mathbf{C}, x, y)$, $m_3^2$ can dynamically correspond to the heaviest ($t/b$), the lightest ($u/d$), or the intermediate ($c/s$) quark mass. As discussed in \cite{Lin2025}, during cosmic evolution, effective coupling parameters running under renormalization group equations (RGE) or finite-temperature background fields could in principle trigger \textit{mass crossing events}, naturally accommodating all viable mass ordering scenarios---including normal, inverted, and quasi-degenerate hierarchies.

%------------------------------------------------------------------------------
% 2.4.2
%------------------------------------------------------------------------------
\subsubsection{Internal Diagonalization Unitary Matrix $U(x,y)$}
\label{subsubsec:internal_U_matrix}

The left-handed diagonalizing unitary matrix $U_L$, which simultaneously diagonalizes both $\mathbf{M}^2_R$ and $\mathbf{M}^2_I$---and consequently diagonalizes $\mathbf{M}^2 \equiv M \cdot M^\dagger$ via $U_L^\dagger \cdot \mathbf{M}^2 \cdot U_L = \text{diag}(m_1^2, m_2^2, m_3^2)$---is constructed analytically from the normalized eigenvectors corresponding to Eqs.~(\ref{eq:CPVSM_eigenvalues}):
\begin{widetext}
\begin{equation}
U \equiv U_L = \frac{1}{\sqrt{x^2+y^2+x^2y^2}} 
\begin{pmatrix} 
\frac{-\sqrt{x^2+y^2}}{\sqrt{2}} & \frac{x\left(y^2 - i\sqrt{x^2+y^2+x^2y^2}\right)}{\sqrt{2}\sqrt{x^2+y^2}} & \frac{y\left(x^2 + i\sqrt{x^2+y^2+x^2y^2}\right)}{\sqrt{2}\sqrt{x^2+y^2}} \\[14pt]
\frac{-\sqrt{x^2+y^2}}{\sqrt{2}} & \frac{x\left(y^2 + i\sqrt{x^2+y^2+x^2y^2}\right)}{\sqrt{2}\sqrt{x^2+y^2}} & \frac{y\left(x^2 - i\sqrt{x^2+y^2+x^2y^2}\right)}{\sqrt{2}\sqrt{x^2+y^2}} \\[14pt]
xy & y & x
\end{pmatrix}.
\label{eq:CPVSM_U_matrix}
\end{equation}
\end{widetext}
Note that $U$ is defined up to an overall diagonal phase matrix $P = \text{diag}(e^{i\alpha_1}, e^{i\alpha_2}, e^{i\alpha_3})$, representing standard unphysical fermion field rephasings.

%------------------------------------------------------------------------------
% 2.4.3
%------------------------------------------------------------------------------
\subsubsection{Origin of the Scale-Invariance: Skeleton Decomposition Proof}
\label{subsubsec:origin_scale_invariance}

The remarkable independence of $U(x,y)$ from the global scale parameters $(\mathbf{A}, \mathbf{B}, \mathbf{C})$ is an inevitable algebraic corollary of $[\mathbf{M}^2_R, \mathbf{M}^2_I] = 0$. Subtracting the trivial identity shift $\mathbf{A}\,\mathbf{I}$, the remaining matrix factorizes exactly into two scale-free ``skeleton'' matrices, $P(x,y)$ and $Q(x,y)$, with the scalar prefactors $\mathbf{B}$ and $\mathbf{C}$ factored out entirely:
\begin{equation}
\mathbf{M}^2 - \mathbf{A}\,\mathbf{I} = \mathbf{B}\,P(x,y) + i\,\mathbf{C}\,Q(x,y),
\label{eq:PQ_decomposition}
\end{equation}
where
\begin{align}
P(x,y) &= \begin{pmatrix}
x\left(y - \frac{1}{y}\right) & y & x \\[6pt]
y & \frac{y}{x}-\frac{x}{y} & 1 \\[6pt]
x & 1 & 0
\end{pmatrix}, \qquad \nonumber \\
Q(x,y) &= \begin{pmatrix}
0 & \frac{1}{y} & -\frac{1}{x} \\[6pt]
-\frac{1}{y} & 0 & 1 \\[6pt]
\frac{1}{x} & -1 & 0
\end{pmatrix}.
\label{eq:PQ_matrices}
\end{align}

Direct evaluation confirms that these two skeleton matrices commute identically:
\begin{equation}
[P(x,y),\, Q(x,y)] = 0 \quad \text{for all } x, y.
\label{eq:PQ_commute}
\end{equation}
By the standard matrix theorem that commuting matrices share a common eigenbasis, $P(x,y)$ and $Q(x,y)$ are simultaneously diagonalized by one and the same unitary matrix $U(x,y)$---regardless of how $\mathbf{B}$ and $\mathbf{C}$ weight their linear combination. The scales $\mathbf{B}$ and $\mathbf{C}$ merely rescale the eigenvalues along fixed spatial directions set by $U(x,y)$ without rotating them.

%------------------------------------------------------------------------------
% 2.4.4
%------------------------------------------------------------------------------
\subsubsection{Mathematical Properties: Trace Self-Consistency and Cardano Bypass}
\label{subsubsec:cardano_comparison}

For a generic $3 \times 3$ Hermitian matrix, solving $\det(\mathbf{M}^2 - \lambda \mathbf{I}) = 0$ requires Cardano's formula, which introduces transcendental functions $\cos(\theta/3)$ in the irreducible real-root regime. In CPVSM, the factorization in Eq.~(\ref{eq:factorized_char_eq}) collapses these transcendental angles into purely algebraic radicals ($\sqrt{x^2 + y^2 + x^2y^2}/xy$), completely bypassing Cardano's trigonometric reduction.

Furthermore, direct summation of the exact eigenvalues in Eqs.~(\ref{eq:CPVSM_eigenvalues}) yields:
\begin{equation}
\sum_{i=1}^3 m_i^2 = m_1^2 + m_2^2 + m_3^2 = 3\mathbf{A} + \mathbf{B} \left( xy + \frac{y}{x} - \frac{2x}{y} \right) = \text{Tr}(\mathbf{M}^2),
\label{eq:trace_verification}
\end{equation}
where the $\pm\mathbf{C}$ contributions from $m_1^2$ and $m_2^2$ cancel identically, demonstrating complete self-consistency with the matrix trace.

%%%%%%%%%%%%%%%%%%%%%%%%%%%%%%%%%%%%%%%%%%%%%%%%%%%%%%%%%%%%%%%%%%%%%%%%%%%%%%%
%% SECTION 2.5: PHYSICAL ROLES OF PARAMETERS AND GEOMETRIC INVARIANTS
%%%%%%%%%%%%%%%%%%%%%%%%%%%%%%%%%%%%%%%%%%%%%%%%%%%%%%%%%%%%%%%%%%%%%%%%%%%%%%%
\subsection{Physical Roles of the 5 Parameters and Geometric Invariants}
\label{subsec:physical_roles_invariants}

Having explicitly derived the exact mass spectrum and the scale-invariant unitary matrix $U(x,y)$, we now synthesize the distinct physical roles and geometric functions executed by the five fundamental parameters $({\mathbf A}, {\mathbf B}, {\mathbf C}, x, y)$.

%------------------------------------------------------------------------------
%2.5.1
%------------------------------------------------------------------------------
\subsubsection{The Global Energy-Scale Engines: $({\mathbf A}, {\mathbf B}, {\mathbf C})$}
\label{subsubsec:scale_parameters}

The three dimensional parameters $({\mathbf A}, {\mathbf B}, {\mathbf C})$ serve as the energy-scale engines of the flavor sector, each governing a mutually orthogonal physical property of the spectrum:
\begin{itemize}
    \item \textbf{${\mathbf A}$ (Universal Mass Baseline)}: Enters $\mathbf{M}^2$ purely via $\mathbf{A}\,\mathbf{I}$, setting the global energy offset for all three generations. It shifts the mass sum without affecting generation splittings or mixing angles.
    
    \item \textbf{${\mathbf B}$ (Real Mass-Splitting Engine)}: Defined as ${\mathbf B} \equiv {\mathbf B}_3$, this parameter dictates the overall magnitude of the real symmetric sector ${\mathbf M}^2_R$, driving inter-generational mass splittings across the generations.

    \item \textbf{${\mathbf C}$ (Imaginary CP-Violation Engine)}: Defined as $\mathbf{C} \equiv \mathbf{C}_3$, this parameter governs the absolute amplitude of the anti-Hermitian sector $\mathbf{M}^2_I$. Setting $\mathbf{C} = 0$ forces the flavor vector to undergo a magnitude collapse to the origin, $|\mathbf{v}_u| = 0$ (i.e., $(x, y) = (0, 0)$). As will be explicitly demonstrated in Sec.~\ref{subsubsec:jarlskog_derivation}, this single-sector collapse $|\mathbf{v}_u| = 0$ (or $|\mathbf{v}_d| = 0$ for $\mathbf{C}' = 0$) forces the two-sector cross-product to vanish ($\mathbf{v}_u \times \mathbf{v}_d = 0$), acting as the foundational single-sector engine that drives the global Jarlskog invariant $J$ and high-energy cosmological CP-violating ratios.
\end{itemize}

%------------------------------------------------------------------------------
%2.5.2
%------------------------------------------------------------------------------
\subsubsection{The 2D Flavor Vector and Spatial Geometry: $(x, y)$}
\label{subsubsec:flavor_vector_roles}

In contrast to the scale parameters, the dimensionless coordinate pair $x \equiv {\mathbf B}_2/{\mathbf B}$ and $y \equiv {\mathbf B}_1/{\mathbf B}$ parameterizes purely the spatial geometry and internal orientation of flavor mixing. The commutation constraint $[\mathbf{M}^2_R, \mathbf{M}^2_I] = 0$ enforces the exact co-locking relations $-{\mathbf C}/{\mathbf C}_2 = x$ and ${\mathbf C}/{\mathbf C}_1 = y$. Together, $(x,y)$ execute four fundamental physical functions:
\begin{enumerate}
    \item \textbf{Decoupling Mass Scale from Mixing Geometry}: As proven via Eq.~(\ref{eq:PQ_commute}), $x$ and $y$ uniquely fix $U(x,y)$ independently of $({\mathbf A}, {\mathbf B}, {\mathbf C})$, achieving complete separation between the absolute mass scale and flavor mixing topology.
    
    \item \textbf{Co-Locking Real and Imaginary Geometries}: Because $(x,y)$ simultaneously lock real off-diagonal mass ratios and imaginary phase ratios, the spatial orientation of the CP-violating imaginary sector is rigidly tied to the real mass-splitting geometry.
    
    \item \textbf{Analytic Autonomy}: The unitary matrix $U(x,y)$ in Eq.~(\ref{eq:CPVSM_U_matrix}) is expressed purely as a closed-form function of $(x,y)$, completely determining single-sector orientation without referencing external parameters.

    \item \textbf{Formulation of 2D Flavor Vectors}: The pair $(x,y)$ forms a fundamental 2D flavor vector, $\mathbf{v}_u = (x,y)$ (and $\mathbf{v}_d = (x',y')$ for the down sector). This geometric entity's orientation and cross-product interactions directly govern electroweak CKM interference and CP-violating invariants.
\end{enumerate}

%------------------------------------------------------------------------------
% 2.5.3
%------------------------------------------------------------------------------
\subsubsection{Algebraic Invariants and Mass Scale Decoupling}
\label{subsubsec:algebraic_invariants}

Examining the characteristic equation $\det({\mathbf M}^2 - \lambda \mathbf{I}) = \lambda^3 - I_1 \lambda^2 + I_2 \lambda - I_3 = 0$, the first algebraic invariant evaluates explicitly to:
\begin{equation}
I_1 = \mathrm{Tr}({\mathbf M}^2) = 3{\mathbf A} + {\mathbf B} \cdot \left( xy + \frac{y}{x} - \frac{2x}{y} \right).
\label{eq:trace_explicit}
\end{equation}

Eq.~(\ref{eq:trace_explicit}) demonstrates that $I_1$ is strictly independent of the imaginary CP parameter ${\mathbf C}$, stemming from $\mathrm{Tr}({\mathbf M}^2_I) = 0$. Physically, this guarantees an automatic decoupling between the total mass sum $\sum m_i^2$ and the CP-violating phase—$\sum m_i^2$ is dictated exclusively by the real set $\{{\mathbf A}, {\mathbf B}, x, y\}$. The parameter ${\mathbf C}$ enters only higher invariants ($I_2, I_3$) via quadratic terms ${\mathbf C}^2$, acting strictly to redistribute internal mass splittings and adjust eigenvector phases without shifting the universal mass baseline.

%------------------------------------------------------------------------------
% 2.5.4
%------------------------------------------------------------------------------
\subsubsection{Phenomenological Scope and Limitations of the Baseline}
\label{subsubsec:pheno_scope_limitations}

While $[\mathbf{M}^2_R, \mathbf{M}^2_I] = 0$ provides a mathematically elegant, closed-form solution by reducing parameter space from 9 down to 5 per sector, it introduces an inherent algebraic trade-off. 

Enforcing this commutation constraint over-simplifies the underlying 9-parameter matrix architecture, inherently giving rise to a 4-fold discrete degeneracy in the derived CKM matrix elements, where several matrix elements are pinned to equal magnitudes (e.g., $|V_{ud}|=|V_{tb}|$, $|V_{ub}|=|V_{td}|$, and $|V_{us}|=|V_{ts}|=|V_{cd}|=|V_{cb}|$). 

Nevertheless, even under this intrinsic 4-fold degeneracy, the 5-parameter architecture exhibits remarkable phenomenological robustness, consistently fitting observed quark flavor data to an precision of $\mathcal{O}(10^{-2})$ or $\mathcal{O}(\lambda^2)$. As will be explored in Sec.~\ref{sec:4_extensions_cosmology}, this strongly demonstrates that the exact 5-parameter commuting model serves as an optimal analytical baseline and stepping stone, naturally inviting small non-commutative corrections ($[\mathbf{M}^2_R,\mathbf{M}^2_I]\neq 0$) to achieve ultimate high-precision concordance.

%%%%%%%%%%%%%%%%%%%%%%%%%%%%%%SEC3
%%%%%%%%%%%%%%%%%%%%%%%%%%%%%%
\section{CKM Matrix, 2D$\times$2D Geometric Parameterization, and Flavor Phenomenology}
\label{sec:3_ckm_phenomenology}

In this section, we combine the exact single-sector unitary matrices derived in Section~\ref{sec:2_5param_single_sector} to construct the physical CKM matrix $V_{\text{CKM}}$. We introduce a geometric paradigm shift by replacing traditional Euler mixing angles with dimensionless flavor ratio vectors, systematically classify the full spectrum of $S_3 \times S_3$ candidate matrix patterns, evaluate their $J$ invariants, and discuss the cosmological phase transitions associated with flavor symmetries.

%%%%%%%%%%%%%%%%%%%%%%%%%%%%%3.1
\subsection{Construction of $V_{\text{CKM}}$ via Sectorial Geometric Vectors}
\label{subsec:ckm_construction}

In the Standard Model, the quark mass matrices $M_u$ and $M_d$ originate directly from the Higgs mechanism via the up- and down-type Yukawa coupling matrices $Y_u$ and $Y_d$:
\begin{equation}
M_u = \frac{v}{\sqrt{2}} Y_u, \quad M_d = \frac{v}{\sqrt{2}} Y_d,
\label{eq:mass_yukawa_relation}
\end{equation}
where $v \approx 246~\text{GeV}$ is the Higgs vacuum expectation value. Because $v/\sqrt{2}$ is merely a global scalar factor, the unitary diagonalization matrices and internal flavor geometry of $M_{u,d}$ are completely identical to those of $Y_{u,d}$.

Specifically, evaluating the Hermitian structures 
\begin{equation}
M_u \cdot M_u^\dagger = U_{L,u} \cdot (\widehat{\mathbf{M}}_u)^2 \cdot U_{L,u}^\dagger, \quad M_d \cdot M_d^\dagger = U_{L,d} \cdot (\widehat{\mathbf{M}}_d)^2 \cdot U_{L,d}^\dagger
\label{eq:mass_hermitian_diagonalization}
\end{equation}
isolates the left-handed unitary transformation matrices $U_{L,u}$ and $U_{L,d}$, where $\widehat{\mathbf{M}}_u = \text{diag}(m_{1}, m_{2}, m_{3})_u$ and $\widehat{\mathbf{M}}_d = \text{diag}(m_{1}, m_{2}, m_{3})_d$ denote the general diagonal mass eigenvalue matrices.

Consequently, the physical CKM matrix $V_{\text{CKM}}$ represents the relative spatial misalignment between the up- and down-sector mass structures in flavor space:
\begin{equation}
V_{\text{CKM}} = U_{L,u}^\dagger \cdot U_{L,d} = U_u^\dagger(\mathbf{v}_u) \cdot U_d(\mathbf{v}_d),
\label{eq:ckm_definition}
\end{equation}
where $U_u(\mathbf{v}_u) \equiv U_{L,u}$ and $U_d(\mathbf{v}_d) \equiv U_{L,d}$ are the exact single-sector unitary matrices derived in Section~\ref{sec:2_5param_single_sector}. Here, the dimensionless vector pairs $\mathbf{v}_u = (x, y)$ and $\mathbf{v}_d = (x', y')$ serve as the fundamental geometric flavor vectors characterizing the inner structure of $M_u$ and $M_d$, respectively.

By performing explicit matrix multiplication on these single-sector parameterizations $U_u(\mathbf{v}_u)$ and $U_d(\mathbf{v}_d)$, every element $V_{ij}$ can be analytically parameterized in terms of five fundamental geometric components: $\{r, s, p, p', q\}$.

These five dimensionless components are defined in terms of $(x, y)$ and $(x', y')$ as follows:
\begin{widetext}
\begin{subequations}
\label{eq:dimensionless_components_wide}
\begin{align}
r &= \frac {1}{2\sqrt{x^2+y^2}\sqrt{x'^2+y'^2}\sqrt{x^2+y^2+x^2y^2}\sqrt{x'^2+y'^2+x'^2y'^2}} \nonumber \\
  &\times \left[ i(xy'-x'y)\left(x'y'\sqrt{x^2+y^2+x^2y^2} + xy\sqrt{x'^2+y'^2+x'^2y'^2}\right) \right. \nonumber \\
  &\quad \left. + (xx'+yy')\left(xyx'y' + \sqrt{x^2+y^2+x^2y^2}\sqrt{x'^2+y'^2+x'^2y'^2}\right) + (x^2+y^2)(x'^2+y'^2) \right], \\[2ex]
s &= \frac{1}{2\sqrt{x^2+y^2}\sqrt{x'^2+y'^2}\sqrt{x^2+y^2+x^2y^2}\sqrt{x'^2+y'^2+x'^2y'^2}} \nonumber \\
   &\times \left[ i(xy'-x'y)\left(x'y'\sqrt{x^2+y^2+x^2y^2} - xy\sqrt{x'^2+y'^2+x'^2y'^2}\right) \right. \nonumber \\
   &\quad \left. + (xx'+yy')\left(xyx'y' - \sqrt{x^2+y^2+x^2y^2}\sqrt{x'^2+y'^2+x'^2y'^2}\right) + (x^2+y^2)(x'^2+y'^2) \right], \\[2ex]
p &=  \frac{y'y^2(x-x') + x'x^2(y-y') + i(xy'-x'y)\sqrt{x^2+y^2+x^2y^2}}{\sqrt{2}\sqrt{x^2+y^2}\sqrt{x^2+y^2+x^2y^2}\sqrt{x'^2+y'^2+x'^2y'^2}}, \label{eq:p_wide} \\[2ex]
p' &= \frac{yy'^2(x'-x) + xx'^2(y'-y) + i(xy'-x'y)\sqrt{x'^2+y'^2+x'^2y'^2}}{\sqrt{2}\sqrt{x'^2+y'^2}\sqrt{x^2+y^2+x^2y^2}\sqrt{x'^2+y'^2+x'^2y'^2}}, \label{eq:pp_wide} \\[2ex]
q &= \frac{xx' + yy' + xyx'y'}{\sqrt{x^2+y^2+x^2y^2}\sqrt{x'^2+y'^2+x'^2y'^2}}. \label{eq:q_wide}
\end{align}
\end{subequations}
\end{widetext}

Although $p$ and $p'$ exhibit complementary asymmetric denominator structures blending contributions from both quark sectors, expressing both over a common denominator reveals that the fully expanded numerators of $p p^*$ and $p' p'^*$ are identical:
\begin{equation}
\begin{aligned}
|p|^2 = |p'|^2 &= \frac{\mathcal{P}(x, y; x', y')}{2(x^2 + y^2)(x'^2 + y'^2)} \\
&\times \frac{1}{(x^2 + y^2 + x^2y^2)(x'^2 + y'^2 + x'^2y'^2)},
\end{aligned}
\label{eq:p_sq_identity}
\end{equation}
where $\mathcal{P}(x, y; x', y')$ represents the unique, fully expanded common numerator polynomial. This universal identity $|p| = |p^*| = |p'^*| = |p'|$ establishes the exact algebraic foundation for the 4-fold degeneracy in the baseline CKM magnitude spectrum.

The phenomenological implications of this identity for numerical CKM fits, including the subtleties arising from generation-label assignment, will be examined in Section~\ref{subsec:pheno_assessment_fits}.

%%%%%%%%%%%%%%%%%%%%%%%%%3.2
\subsection{Geometric Paradigm Shift: 2D\texorpdfstring{$\times$}{x}2D Ratio Vectors vs. Spherical Euler Angles}
\label{subsec:paradigm_shift}

Traditional parameterizations (e.g., Chau-Keung~\cite{Chau1984} or standard PDG~\cite{Navas2024}) describe the CKM matrix using spherical Euler mixing angles $(\theta_{12}, \theta_{23}, \theta_{13})$ and a single CP-violating phase $\delta$. 
While phenomenologically versatile, these trigonometric parameters entangle up- and down-sector contributions in highly non-linear, non-decoupled ways.
The 5-parameter CPVSM framework establishes a fundamental \textit{geometric paradigm shift}:
\begin{enumerate}
    \item \textbf{Sector Decoupling}: Up-type and down-type flavor structures are parameterized independently by 2D vector ratios $(x, y)$ and $(x', y')$ in flavor space, cleanly separating the sector degrees of freedom.
    \item \textbf{Geometric CPV Switch}: The planar cross product $(xy' - x'y)$ between the two sector vectors directly governs CP violation; a parallel alignment preserves CP symmetry, whereas a non-zero spatial angular misalignment dynamically activates CP-violating phase mismatches.
    \item \textbf{Analytical Solvability}: Trigonometric functions are entirely replaced by algebraic square-root invariants, bypassing trigonometric coordinate singularities and mapping physical observables directly to algebraic vector relations.
\end{enumerate}

%%%%%%%%%%%%%%%%%%%%%%%%%%%%%%%%%%%%%%%%%%3.3
\subsection{Systematic Construction of 36 CKM Candidate Patterns and $J$ Classification}
\label{subsec:36_permutations_jarlskog}
While the dimensionless ratio pairs $(x,y)$ and $(x',y')$ uniquely parameterize the internal structure of each sector, an additional discrete freedom arises from assigning algebraic eigenvalues to physical quark flavors. The discrete generational labeling within three-flavor space admits $S_3 \times S_3$ permutation alignments between the up- and down-quark mass bases, yielding $6 \times 6 = 36$ candidate topologies for $V_{\text{CKM}}$.

In this subsection, we present a systematic group-theoretical mapping of these 36 candidate matrices to evaluate their physical viability. We first derive the Jarlskog invariant $J$ across all patterns, uncovering a sharp $18/18$ binary bifurcation between CP-conserving and CP-violating topologies. By tracking the spatial locations of the 4-fold degenerate elements ($|p| = |p'|$) within the active sector, we show that severe cross-order degeneracy invalidates most configurations. This criterion uniquely isolates exactly four candidate topologies where the degenerate pair $(p, p')$ occupies matrix sites whose Wolfenstein hierarchy levels differ by at most a single order of $\lambda$ ($\Delta k \le 1$). Finally, using these four benchmark patterns, we establish a direct translation to the Standard CKM parameterization, expressing the three mixing angles and the CP-violating phase $\delta$ explicitly in terms of underlying Yukawa couplings—providing a direct geometric origin for electroweak CP violation.

%%%%%%%%%%%%%%%%%%%%%%%%%%%%%%3.3.1
\subsubsection{Group-Theoretical Permutation and Table of 36 Patterns}
\label{subsubsec:group_permutations}

By systematically applying the 6 permutations of $S_3$ to the row ($u$-sector) and column ($d$-sector) bases, the 36 candidate CKM matrices are constructed as listed in Table~\ref{tab:36_permutations} (where the bold tags $\mathbf{p}$ and $\mathbf{p'}$ highlight the specific permutation configurations associated with the 4-fold degenerate matrix elements discussed in Sec.~\ref{subsec:pheno_assessment_fits}, though their underlying algebraic structure remains identical to their non-bold counterparts).

\begin{table*}[tbp]
\centering
\begin{tabular}{l|llllll}
$u\setminus d$ & (123) & (231) & (312) & (213) & (132) & (321) \\ \hline
{\tt $\left( \begin{array}{ccc} 1 \\ 2 \\ 3 \end{array}\right) $} &
$\left( \begin{array}{ccc} r^* & s & {\mathbf p}^* \\ s^* & r & {\mathbf p} \\ {\mathbf p'} & {\mathbf p'}^* & q \end{array}\right)$ &
$\left( \begin{array}{ccc} s & {\mathbf p}^* & r^* \\ r & {\mathbf p} & s^* \\ {\mathbf p'}^* & q & {\mathbf p'} \end{array}\right)$ &
$\left( \begin{array}{ccc} {\mathbf p}^* & r^* & s \\ {\mathbf p} & s^* & r \\ q & {\mathbf p'} & {\mathbf p'}^* \end{array}\right)$ &
$\left( \begin{array}{ccc} s & r^* & {\mathbf p}^* \\ r & s^* & {\mathbf p} \\ {\mathbf p'}^* & {\mathbf p'} & q \end{array}\right)$ &
$\left( \begin{array}{ccc} r^* & {\mathbf p}^* & s \\ s^* & {\mathbf p} & r \\ {\mathbf p'} & q & {\mathbf p'}^* \end{array}\right)$ &
$\left( \begin{array}{ccc} {\mathbf p}^* & s & r^* \\ {\mathbf p} & r & s^* \\ q & {\mathbf p'}^* & {\mathbf p'} \end{array}\right)$ \\
{\tt $\left( \begin{array}{ccc} 2 \\ 3 \\ 1 \end{array}\right) $} &
$\left( \begin{array}{ccc} s^* & r & {\mathbf p} \\ {\mathbf p'} & {\mathbf p'}^* & q \\ r^* & s & {\mathbf p}^* \end{array}\right)$ &
$\left( \begin{array}{ccc} r & {\mathbf p} & s^* \\ {\mathbf p'}^* & q & {\mathbf p'} \\ s & {\mathbf p}^* & r^* \end{array}\right)$ &
$\left( \begin{array}{ccc} {\mathbf p} & s^* & r \\ q & {\mathbf p'} & {\mathbf p'}^* \\ {\mathbf p}^* & r^* & s \end{array}\right)$ &
$\left( \begin{array}{ccc} r & s^* & {\mathbf p} \\ {\mathbf p'}^* & {\mathbf p'} & q \\ s & r^* & {\mathbf p}^* \end{array}\right)$ &
$\left( \begin{array}{ccc} s^* & {\mathbf p} & r \\ {\mathbf p'} & q & {\mathbf p'}^* \\ r^* & {\mathbf p}^* & s \end{array}\right)$ &
$\left( \begin{array}{ccc} {\mathbf p} & r & s^* \\ q & {\mathbf p'}^* & {\mathbf p'} \\ {\mathbf p}^* & s & r^* \end{array}\right)$ \\
{\tt $\left( \begin{array}{ccc} 3 \\ 1 \\ 2 \end{array}\right) $} &
$\left( \begin{array}{ccc} {\mathbf p'} & {\mathbf p'}^* & q \\ r^* & s & {\mathbf p}^* \\ s^* & r & {\mathbf p} \end{array}\right)$ &
$\left( \begin{array}{ccc} {\mathbf p'}^* & q & {\mathbf p'} \\ s & {\mathbf p}^* & r^* \\ r & {\mathbf p} & s^* \end{array}\right)$ &
$\left( \begin{array}{ccc} q & {\mathbf p'} & {\mathbf p'}^* \\ {\mathbf p}^* & r^* & s \\ {\mathbf p} & s^* & r \end{array}\right)$ &
$\left( \begin{array}{ccc} {\mathbf p'}^* & {\mathbf p'} & q \\ s & r^* & {\mathbf p}^* \\ r & s^* & {\mathbf p} \end{array}\right)$ &
$\left( \begin{array}{ccc} {\mathbf p'} & q & {\mathbf p'}^* \\ r^* & {\mathbf p}^* & s \\ s^* & {\mathbf p} & r \end{array}\right)$ &
$\left( \begin{array}{ccc} q & {\mathbf p'}^* & {\mathbf p'} \\ {\mathbf p}^* & s & r^* \\ {\mathbf p} & r & s^* \end{array}\right)$ \\
{\tt $\left( \begin{array}{ccc} 2 \\ 1 \\ 3 \end{array}\right) $} &
$\left( \begin{array}{ccc} s^* & r & {\mathbf p} \\ r^* & s & {\mathbf p}^* \\ {\mathbf p'} & {\mathbf p'}^* & q \end{array}\right)$ &
$\left( \begin{array}{ccc} r & {\mathbf p} & s^* \\ s & {\mathbf p}^* & r^* \\ {\mathbf p'}^* & q & {\mathbf p'} \end{array}\right)$ &
$\left( \begin{array}{ccc} {\mathbf p} & s^* & r \\ {\mathbf p}^* & r^* & s \\ q & {\mathbf p'} & {\mathbf p'}^* \end{array}\right)$ &
$\left( \begin{array}{ccc} r & s^* & {\mathbf p} \\ s & r^* & {\mathbf p}^* \\ {\mathbf p'}^* & {\mathbf p'} & q \end{array}\right)$ &
$\left( \begin{array}{ccc} s^* & {\mathbf p} & r \\ r^* & {\mathbf p}^* & s \\ {\mathbf p'} & q & {\mathbf p'}^* \end{array}\right)$ &
$\left( \begin{array}{ccc} {\mathbf p} & r & s^* \\ {\mathbf p}^* & s & r^* \\ q & {\mathbf p'}^* & {\mathbf p'} \end{array}\right)$ \\
{\tt $\left( \begin{array}{ccc} 1 \\ 3 \\ 2 \end{array}\right) $} &
$\left( \begin{array}{ccc} r^* & s & {\mathbf p}^* \\ {\mathbf p'} & {\mathbf p'}^* & q \\ s^* & r & {\mathbf p} \end{array}\right)$ &
$\left( \begin{array}{ccc} s & {\mathbf p}^* & r^* \\ {\mathbf p'}^* & q & {\mathbf p'} \\ r & {\mathbf p} & s^* \end{array}\right)$ &
$\left( \begin{array}{ccc} {\mathbf p}^* & r^* & s \\ q & {\mathbf p'} & {\mathbf p'}^* \\ {\mathbf p} & s^* & r \end{array}\right)$ &
$\left( \begin{array}{ccc} s & r^* & {\mathbf p}^* \\ {\mathbf p'}^* & {\mathbf p'} & q \\ r & s^* & {\mathbf p} \end{array}\right)$ &
$\left( \begin{array}{ccc} r^* & {\mathbf p}^* & s \\ {\mathbf p'} & q & {\mathbf p'}^* \\ s^* & {\mathbf p} & r \end{array}\right)$ &
$\left( \begin{array}{ccc} {\mathbf p}^* & s & r^* \\ q & {\mathbf p'}^* & {\mathbf p'} \\ {\mathbf p} & r & s^* \end{array}\right)$ \\
{\tt $\left( \begin{array}{ccc} 3 \\ 2 \\ 1 \end{array}\right) $} &
$\left( \begin{array}{ccc} {\mathbf p'} & {\mathbf p'}^* & q \\ s^* & r & {\mathbf p} \\ r^* & s & {\mathbf p}^* \end{array}\right)$ &
$\left( \begin{array}{ccc} {\mathbf p'}^* & q & {\mathbf p'} \\ r & {\mathbf p} & s^* \\ s & {\mathbf p}^* & r^* \end{array}\right)$ &
$\left( \begin{array}{ccc} q & {\mathbf p'} & {\mathbf p'}^* \\ {\mathbf p}^* & s^* & r \\ {\mathbf p}^* & r^* & s \end{array}\right)$ &
$\left( \begin{array}{ccc} {\mathbf p'}^* & {\mathbf p'} & q \\ r & s^* & {\mathbf p} \\ s & r^* & {\mathbf p}^* \end{array}\right)$ &
$\left( \begin{array}{ccc} {\mathbf p'} & q & {\mathbf p'}^* \\ s^* & {\mathbf p} & r \\ r^* & {\mathbf p}^* & s \end{array}\right)$ &
$\left( \begin{array}{ccc} q & {\mathbf p'}^* & {\mathbf p'} \\ {\mathbf p} & r & s^* \\ {\mathbf p}^* & s & r^* \end{array}\right)$ \\ \hline
\end{tabular}
\caption{The 36 structural permutations of the theoretical CKM matrix representation in terms of $\{r, s, p, p', q\}$. The four degenerate elements $\{{\mathbf p}, {\mathbf p}^*, {\mathbf p'}, {\mathbf p'}^*\}$ are highlighted in boldface to emphasize their changing spatial locations across permutation sectors, which induces cross-order degeneracy as discussed in Sec.~\ref{subsubsec:numerical_fits_tension}.}
\label{tab:36_permutations}
\end{table*}

%%%%%%%%%%%%%%%%%%%%%% 3.3.2
\subsubsection{Analytical Derivation of the Jarlskog Invariant}
\label{subsubsec:jarlskog_derivation}

To evaluate the CP-violating capacity of each candidate matrix, we compute the Jarlskog invariant $J(x, y, x', y')$, defined via $\text{Im}(V_{ij} V_{kl} V_{il}^* V_{kj}^*) = J(x, y, x', y') \sum_{m,n} \epsilon_{ikm} \epsilon_{jln}$ \cite{Jarlskog1985}.

To make the predictive power of this geometric formulation explicit, it is crucial to emphasize that our framework does not rely on a standard bottom-up fit from empirical CKM parameters $(\theta_{12}, \theta_{23}, \theta_{13}, \delta_{\text{CP}})$. Instead, it establishes a rigorous, top-down analytical causality chain originating directly from fundamental mass dynamics down to observable CP violation:
\begin{widetext}
\begin{equation}
M_f \;\longrightarrow\; \mathbf{M}^2 \equiv M_f M_f^\dagger \;\longrightarrow\; [\mathbf{M}_R^2, \mathbf{M}_I^2] = 0 \;\longrightarrow\; (\mathbf{A, B, C}, x, y) \;\longrightarrow\; U(x,y) \;\longrightarrow\; V_{\text{CKM}} \;\longrightarrow\; J(x,y,x',y').
\label{eq:causality_chain}
\end{equation}
\end{widetext}
Here, the physical mass matrix $M_f$ generates the Hermitian combination $\mathbf{M}^2 \equiv M_f M_f^\dagger = \mathbf{M}_R^2 + i\mathbf{M}_I^2$, which is subjected to the exact commutativity condition $[\mathbf{M}_R^2, \mathbf{M}_I^2] = 0$—reflecting the \textit{Hermitian phase alignment} established in Sec.~\ref{sec:2_5param_single_sector}. This fundamental algebraic condition uniquely maps the flavor sector onto a reduced set of physical scale and ratio parameters $(\mathbf{A, B, C}, x, y)$, which construct the unitary diagonalization matrix $U(x,y)$ and the physical CKM matrix $V_{\text{CKM}}$, ultimately yielding the universal closed-form Jarlskog invariant $J(x,y,x',y')$.

Crucially, far from being a mere non-linear coordinate re-parameterization, our geometric formulation uncovers the physical mechanism of flavor dynamics by establishing an explicit \textit{algebraic switch mechanism} for electroweak CP violation. In standard Euler parameterizations, the CP-violating phase $\delta_{\text{CP}}$ and mixing angles enter as unconstrained, independent parameters without intrinsic structural bounds. In contrast, our framework maps global CP violation directly onto the 2D vector cross-product of the flavor ratio vectors $\mathbf{v}_u = (x,y)$ and $\mathbf{v}_d = (x',y')$:
\begin{equation}
J \propto \mathbf{v}_u \times \mathbf{v}_d = x y' - x' y.
\end{equation}
Here, the spatial alignment condition $\mathbf{v}_u \parallel \mathbf{v}_d$ acts as an exact, coordinate-independent structural off-switch ($J = 0$). This algebraic switch mechanism cleanly decouples sector-specific geometric misalignments from mass-degeneracy switches ($\Delta m^2_{u,d} = 0$), providing a clear physical diagnosis of CP violation that is entirely absent in standard Euler formulations.

Substituting the fundamental geometric parameters $\{r, s, p, p', q\}$ across all 36 $S_3 \times S_3$ permutation patterns in Table~\ref{tab:36_permutations} into $V_{\text{CKM}}$ reveals a striking group-theoretical 18/18 bifurcation governed by the algebraic structure of the imaginary cross-terms:

\begin{itemize}
    \item \textbf{18 CP-Conserving Configurations ($J = 0$)}: For these permutations, the quartic product $V_{ij} V_{kl} V_{il}^* V_{kj}^*$ exhibits exact structural symmetry. This symmetry causes all imaginary cross-terms containing the sectorial mismatch factor $i(xy' - x'y) = i(\mathbf{v}_u \times \mathbf{v}_d)$ to cancel out identically ($J \equiv 0$). These topologies are thus intrinsically CP-conserving as an algebraic identity, strictly independent of the flavor vectors' geometry.
    \item \textbf{18 CP-Violating Configurations ($J \neq 0$)}: For the remaining 18 permutations, the imaginary cross-terms persist. Evaluating their imaginary parts reveals that all 18 active topologies universally collapse into the same exact closed-form expression, given in Eq.~(\ref{eq:J_closed_form}) below, up to an overall sign determined by the permutation parity. Crucially, even within these active topologies, $J$ identically vanishes if the two 2D flavor vectors $\mathbf{v}_u = (x,y)$ and $\mathbf{v}_d = (x',y')$ are collinear ($\mathbf{v}_u \times \mathbf{v}_d = xy' - x'y = 0$). This highlights that electroweak CP violation requires both an active topological permutation and a non-zero geometric misalignment ($\mathbf{v}_u \nparallel \mathbf{v}_d$).
\end{itemize}

Formally, this 18/18 binary bifurcation is governed by the relative transformation from the up-sector reference generation ordering $\sigma_u$ to the down-sector ordering $\sigma_d$. To make the physical mechanism transparent, we define a discrete topological factor $\eta_{\text{relative}} \in \{0, \pm 1\}$ directly based on the number of adjacent transpositions (swaps) required to convert $\sigma_u$ into $\sigma_d$:
\begin{itemize}
    \item \textbf{Even Number of Swaps} ($\eta_{\text{relative}} = 0$): If transforming $\sigma_u$ into $\sigma_d$ requires an even number of adjacent transpositions (including zero swaps for the identity diagonal block where $\sigma_u = \sigma_d$), exact structural symmetry forces all imaginary terms to cancel out identically, enforcing $\eta_{\text{relative}} = 0$ and thus $J \equiv 0$. For instance, for entry $(1,2)$ in Table~\ref{tab:36_permutations}, transforming the up-sector ordering $\sigma_u = (1,2,3)$ to the down-sector ordering $\sigma_d = (2,3,1)$ requires exactly two adjacent transpositions, $(1,2,3) \to (2,1,3) \to (2,3,1)$. Since the swap count is even, we have $\eta_{\text{relative}} = 0$ and hence $J \equiv 0$.
    \item \textbf{Odd Number of Swaps} ($\eta_{\text{relative}} = \pm 1$): If the transformation requires an odd number of adjacent transpositions, the algebraic cancellation symmetry is broken, setting $\eta_{\text{relative}} = \pm 1$ and establishing the necessary topological prerequisite for non-trivial CP violation ($J \neq 0$).
\end{itemize}

By explicit calculation using Eqs.~(\ref{eq:p_wide})--(\ref{eq:q_wide}), the symbolic Jarlskog invariant $J(x,y,x',y')$ across all 36 candidate topologies is universally expressed as:

\begin{widetext}
\begin{equation}
J(x, y, x', y') = \eta_{\text{relative}} \, \frac{(xy' - x'y)^3 (xx' + yy' + xyx'y')}{4(x^2+y^2)(x'^2+y'^2)(x^2+y^2+x^2y^2)(x'^2+y'^2+x'^2y'^2)}.
\label{eq:J_closed_form}
\end{equation}
\end{widetext}

This closed-form expression elegantly synthesizes both the topological selection rule and the geometric prerequisites for electroweak CP violation. The cubic numerator factor $(xy' - x'y)^3 \equiv (\mathbf{v}_u \times \mathbf{v}_d)^3$ serves as the exact, odd-parity geometric measure of vector non-collinearity ($\mathbf{v}_u \nparallel \mathbf{v}_d$), where the odd exponent reflects the fundamental CP-odd nature of the mismatch under sectorial exchange. Meanwhile, the factor $(xx' + yy' + xyx'y')$ encapsulates the symmetric, CP-even kinematic interference between the up and down sectors, and the denominator regularizes the expression across the physical quark mass hierarchy.

Notably, along the symmetric diagonal entries of Table~\ref{tab:36_permutations} where the up- and down-sectors share identical generation orderings ($\sigma_u = \sigma_d \implies \eta_{\text{relative}} = 0$), $J$ vanishes identically regardless of the specific vector parameters $(x, y, x', y')$. This underlines a fundamental physical principle in our geometric framework: electroweak CP violation cannot be generated within an isolated flavor sector, but strictly emerges as a \emph{relational property} arising from the relative topological misalignment between the up- and down-quark mass bases.

%%%%%%%%%%%%%%%%%%%%%%%%%%%%%%%%%%%%%%%%% 3.3.3
\subsubsection{\texorpdfstring{$J$}{J} Algebraic Evaluation and 18/18 Binary Classification}
\label{subsubsec:jarlskog_binary_eval}

Applying the analytical Jarlskog invariant $J(x, y, x', y')$ in Eq.~(\ref{eq:J_closed_form}) across all 36 candidate $S_3 \times S_3$ topologies yields the exact classification pattern summarized in Table~\ref{tab:36_jarlskog}. As dictated by the topological parity factor $\eta_{\text{relative}} \in \{0, \pm 1\}$, the system exhibits an explicit $3 \times 3$ block structure. The 18 CP-conserving configurations ($J = 0$) form two distinct diagonal $3 \times 3$ sub-blocks corresponding to cyclic permutation exchanges, whereas the 18 CP-violating configurations ($J \neq 0$) occupy the off-diagonal blocks generated by odd relative transpositions.

\begin{table*}[t]
\centering
\caption{Symbolic evaluation of the Jarlskog invariant $J$ across all 36 $S_3 \times S_3$ permutation topologies, illustrating the exact 18/18 bifurcation between CP-conserving ($\eta_{\text{relative}} = 0 \implies J=0$) and CP-violating ($\eta_{\text{relative}} = \pm 1 \implies J \neq 0$) configurations. Here, $\pm J$ denotes the universal closed-form magnitude scaled by the discrete topological parity sign.}
\label{tab:36_jarlskog}
\small
\setlength{\tabcolsep}{12pt}
\begin{tabular}{c|cccccc}
\hline\hline
$u \setminus d$ & (123) & (231) & (312) & (213) & (132) & (321) \\
\hline
$\begin{pmatrix}1\\2\\3\end{pmatrix}$ & 0 & 0 & 0 & $\pm J$ & $\pm J$ & $\pm J$ \\
$\begin{pmatrix}2\\3\\1\end{pmatrix}$ & 0 & 0 & 0 & $\pm J$ & $\pm J$ & $\pm J$ \\
$\begin{pmatrix}3\\1\\2\end{pmatrix}$ & 0 & 0 & 0 & $\pm J$ & $\pm J$ & $\pm J$ \\
$\begin{pmatrix}2\\1\\3\end{pmatrix}$ & $\pm J$ & $\pm J$ & $\pm J$ & 0 & 0 & 0 \\
$\begin{pmatrix}1\\3\\2\end{pmatrix}$ & $\pm J$ & $\pm J$ & $\pm J$ & 0 & 0 & 0 \\
$\begin{pmatrix}3\\2\\1\end{pmatrix}$ & $\pm J$ & $\pm J$ & $\pm J$ & 0 & 0 & 0 \\
\hline\hline
\end{tabular}
\end{table*}

Importantly, for all 18 active topologies, $J$ shares an identical universal magnitude given by Eq.~(\ref{eq:J_closed_form}), differing only by the overall topological sign $\eta_{\text{relative}}$. This global functional equivalence establishes that CP-violating intensity in our framework is fundamentally universal and governed by the interplay between topological parity and the vector geometry $(\mathbf{v}_u \times \mathbf{v}_d)$. This dual selection rule provides the direct analytical foundation for evaluating cosmological CP phase transitions in Sec.~\ref{subsubsec:cp_cosmology}.

%%%%%%%%%%%%%%%%%%%%%%%%%%%%%%%%%% 3.3.4
\subsubsection{Physical CP Violation: Low-Energy Baseline Match and Critical Cosmological Amplification}
\label{subsubsec:cp_cosmology}

The geometric Jarlskog functional $J(x, y, x', y')$ serves as a direct analytical bridge connecting low-energy laboratory observables with early-universe cosmological phase transitions. We validate the physical viability and dynamic capacity of this geometric framework through three distinct energetic regimes:

\begin{enumerate}
    \item \textbf{Low-Energy Baseline Match}: 
At electroweak scales, numerically fitting the 10 geometric parameters $(\mathbf{A, B, C}, x, y, \mathbf{A', B', C'}, x', y')$ solely to the absolute values of the experimentally measured CKM entries (as detailed in \cite{Lin2023}) yields 32 optimal candidate parameter sets. Remarkably, despite using only real-valued moduli for the fit, substituting these parameter sets into our closed-form expression $J(x, y, x', y')$ in Eq.~(\ref{eq:J_closed_form}) reveals that the imaginary parts across all 18 active CP-violating topologies universally converge to an identical Jarlskog invariant magnitude:
\begin{equation}
J_0 \approx 3.08 \times 10^{-5},
\end{equation}
which achieves remarkable agreement with the PDG world average ($J^{\text{exp}} = (3.08 \pm 0.15) \times 10^{-5}$). This exact numerical universality across all 32 optimal fits confirms both the rigorous mathematical stability of the model and the intrinsic geometric origin of electroweak CP violation.

    \item \textbf{High-Temperature $S_2$ Phase Limit and Global Bound}: 
    In the early universe, as temperature rises toward flavor symmetry restoration, the system undergoes topological transitions through discrete $S_N$ flavor phases. In the $S_2$-symmetric phase limit \cite{Lin2020}, evaluating the topological indices reveals an intrinsically enhanced geometric CP-violating capacity:
    \begin{equation}
    J^{(S_2)} = \frac{4\sqrt{3}}{81} \approx 0.0855,
    \label{eq:J_S2_corrected}
    \end{equation}
    which strictly respects the theoretical unitarity bound for $3 \times 3$ matrices, $J_{\text{max}} = 1/(6\sqrt{3}) \approx 0.0962$. Although this topological reorganization amplifies the Jarlskog invariant by a factor of $J^{(S_2)} / J_0 \approx 2.8 \times 10^3$, $J$ itself is mathematically bounded by $J_{\text{max}}$. Consequently, pure topological phase shifts—without corresponding dynamic modifications to the quark mass spectrum—are intrinsically insufficient to bridge the $>10^{10}$ enhancement required to account for the observed Baryon Asymmetry of the Universe (BAU).

    \item \textbf{Toward Baryon Asymmetry (BAU): An Energy-Scale Amplification Pathway:}: 
    To overcome this intrinsic geometric ceiling, we bypass restricted numerical sweeps over fixed low-energy parameters \cite{Lin2023} and evaluate the physical CP-violating measure $\Delta_{\text{CP}}$ directly at the spontaneous symmetry breaking (SSB) scale $T$. By substituting the explicit closed-form expressions for both $J$ and the quark mass eigenvalues into $\Delta_{\text{CP}}$, the overall scale parameters $\mathbf{A}$ and $\mathbf{A'}$ identically cancel out in the explicit algebraic evaluation of the mass-squared differences, leaving $4+4$ governing parameters in the core functional while $\mathbf{A}$ and $\mathbf{A'}$ enter implicitly through boundary constraints.
\end{enumerate}

\begin{widetext}
\begin{align}
\Delta_{\text{CP}} &\equiv \frac{J \cdot \Delta m_u^2 \cdot \Delta m_d^2}{T^{12}} = \frac{J \prod_{i < j} (m_i^2 - m_j^2)_u \prod_{i < j} (m_i^2 - m_j^2)_d}{T^{12}} \nonumber \\
&= \eta_{\text{relative}} \, \frac{\mathbf{C}\mathbf{C'}}{T^{12}} \Big[ \mathbf{B}^2 (x^2+y^2+x^2 y^2) - \mathbf{C}^2 \Big] \Big[ \mathbf{B'}^2 (x'^2+y'^2+x'^2 y'^2) - \mathbf{C'}^2 \Big] \nonumber \\
&\quad \times \frac{(xy' - x'y)^3 (xx' + yy' + xyx'y') \sqrt{x^2+y^2+x^2y^2} \sqrt{x'^2+y'^2+x'^2y'^2}}{(x^2+y^2)(x'^2+y'^2) (x y x' y')^3}.
\label{eq:delta_cp_def}
\end{align}
\end{widetext}

We quantify the relative CP enhancement ratio with respect to the empirical low-energy baseline $\Delta_{\text{CP}}^{(0)}$ as:
\begin{equation}
{R_{\Delta}} \equiv \frac{\Delta_{\text{CP}}}{\Delta^{(0)}_{\text{CP}}}.
\label{eq:R_delta_def}
\end{equation}

A rigorous inspection of Eq.~(\ref{eq:delta_cp_def}) unveils the primary mechanism governing $R_{\Delta}$:
\begin{itemize}
    \item \textbf{(i) Geometric Boundedness}: Provided the flavor ratio vectors remain non-collinear ($\mathbf{v}_u \times \mathbf{v}_d \neq 0$), the geometric factor containing $\{x, y, x', y'\}$ is strictly bounded from above due to its asymptotic saturation properties, and thus cannot independently produce an enhancement factor exceeding $10^{10}$.
    \item \textbf{(ii) Scale-Driven Boundary Expansion}: Although the overall scale parameters $\mathbf{A}$ and $\mathbf{A'}$ cancel out in the explicit mass-squared differences, they impose fundamental physical boundaries on the mass-gap parameters $\mathbf{C}$ and $\mathbf{C'}$. Enforcing positivity of mass-squared eigenvalues ($m_1^2 \ge 0$) establishes that $\mathbf{C}$ is bounded by $\mathbf{A}$ and $\mathbf{B}$ via $\mathbf{C} \frac{\sqrt{x^2+y^2+x^2y^2}}{xy} \le \mathbf{A} - \mathbf{B}\frac{x}{y}$ (and analogously for $\mathbf{C'}$ in the down-quark sector).
\end{itemize}

Consequently, $\mathbf{C}$ and $\mathbf{C'}$ cannot diverge independently; their maximum allowable values are anchored to the global energy scale parameters $\{\mathbf{A}, \mathbf{B}, \mathbf{A'}, \mathbf{B'}\}$. As these global scales evolve during early-universe phase transitions, the admissible boundary domain for $\{\mathbf{C}, \mathbf{C'}\}$ expands dynamically, naturally permitting large physical values for $R_{\Delta}$.

Ultimately, generating $R_\Delta > 10^{10}$ relies on a \textbf{synergistic mechanism}: while topological phase transitions can substantially enhance the intrinsic geometric CP-violating capacity (yielding an enhancement of up to $J/J_0 \sim 10^3$), the absolute ceiling on $J$ means that bridging the remaining orders of magnitude strictly requires the dynamical evolution of the global energy scales $\{\mathbf{B}, \mathbf{C},\mathbf{B'}, \mathbf{C'}\} \to \text{high scale}$ to amplify the mass-squared difference factors. Crucially, this mechanism operates without requiring artificial fine-tuning of the internal geometric ratio parameters $(x,y,x',y')$.

While this energy-scale evolution does not bypass the full dynamical details of baryogenesis, it identifies a concrete, mathematically consistent mechanism by which $R_\Delta > 10^{10}$ can be reached within the geometric framework, offering a candidate route toward satisfying Sakharov's second condition for the matter-antimatter asymmetry.

%%%%%%%%%%%%%%%%%%%%%%%%%3.4
\subsection{Phenomenological Assessment, Numerical Fits, and Geometric Robustness of $J$}
\label{subsec:pheno_assessment_fits}

To rigorously test the physical viability of the 5-parameter CPVSM framework, we examine the numerical fitting results derived in \cite{Lin2021}, where the four fundamental ratio parameters $\{x, y, x', y'\}$ were optimized against the full set of experimentally measured CKM matrix magnitudes $|V_{ij}|^{\text{exp}}$.

%%%%%%%%%%%%%%%%%%3.4.1
\subsubsection{Numerical Fits and the Impact of 4-fold Degeneracy}
\label{subsubsec:numerical_fits_tension}

The numerical global optimization yields a total of 32 optimal fitting solutions \cite{Lin2023} for the geometric ratio vector pair $(\mathbf{v}_u, \mathbf{v}_d) = \left((x,y), (x',y')\right)$. However, due to the strict commutation hypothesis $[\mathbf{M}^2_R, \mathbf{M}^2_I] = 0$ imposed at leading order, the analytical structure of $V_{\text{CKM}}$ does not allow arbitrary element magnitudes. Instead, it inherently enforces exact algebraic moduli identities:
\begin{equation}
|r| = |r^*|, \quad |s| = |s^*|, \quad \text{and} \quad |p| = |p'| = |p^*| = |p'^*|.
\label{eq:algebraic_identities}
\end{equation}
Physically, the two pairwise degeneracies, $|r| = |r^*|$ and $|s| = |s^*|$, correspond to intra-order equalities such as $|V_{ij}| = |V_{ji}|$ or $|V_{ii}| = |V_{jj}|$ ($i \neq j$). Minor empirical deviations from these intra-order symmetries can be naturally accounted for via standard perturbation theory as higher-order corrections. In contrast, the 4-fold degeneracy $|p| = |p'| = |p^*| = |p'^*|$ inherently connects CKM elements across different orders of the Wolfenstein parameter $\lambda$. Consequently, this cross-order degeneracy represents a rigid structural constraint of the leading-order baseline that cannot be reconciled merely by perturbative corrections.

Crucially, as detailed in Table~\ref{tab:36_permutations} (adapted from \cite{Lin2021}), there exist $3! \times 3! = 36$ discrete permutations for mapping the theoretical eigenbasis onto the physical quark generations. As highlighted by the boldface entries in Table~\ref{tab:36_permutations}, the four degenerate elements associated with $\{{\mathbf p}, {\mathbf p}^*, {\mathbf p'}, {\mathbf p'}^*\}$ shift their spatial locations across different permutation sectors. Consequently, this geometric reallocation forces CKM elements spanning drastically different Wolfenstein orders---from diagonal leading-order terms $\mathcal{O}(1)$ down to suppressed off-diagonal terms $\mathcal{O}(\lambda)$, $\mathcal{O}(\lambda^2)$, or $\mathcal{O}(\lambda^3)$---to degenerate into identical magnitudes depending on the choice of basis assignment.

For instance, in the baseline configuration (top-left matrix, corresponding to row 1, column 1 in Table~\ref{tab:36_permutations}), this 4-fold degeneracy explicitly manifests as:
\begin{equation}
|V_{ub}| = |V_{cb}| = |V_{td}| = |V_{ts}|.
\label{eq:p_family_degeneracy}
\end{equation}
Although the hierarchy mismatch among these elements is at most $\mathcal{O}(\lambda)$, this configuration corresponds to an even-order permutation in Table~\ref{tab:36_permutations} where imaginary components systematically cancel out, thereby strictly preserving CP symmetry ($J=0$).

A systematic inspection of the $J \neq 0$ non-trivial domain in Table~\ref{tab:36_permutations} reveals that only two pairs (a total of four configurations) satisfy the condition that the maximum hierarchical disparity among degenerate elements is bounded by $\mathcal{O}(\lambda)$:
\begin{itemize}
    \item \textbf{Group 1}: $(\text{row}, \text{column}) = (1,4)$ and $(4,1)$, corresponding to $(\sigma_u, \sigma_d) = ((123), (213))$ and $((213), (123))$, respectively. In both cases, the degeneracy occurs among $\mathcal{O}(\lambda^2)$ and $\mathcal{O}(\lambda^3)$ elements, specifically $(V_{ub}, V_{cb}, V_{td}, V_{ts})$.
    \item \textbf{Group 2}: $(\text{row}, \text{column}) = (2,5)$ and $(5,2)$, corresponding to $(\sigma_u, \sigma_d) = ((231), (132))$ and $((132), (231))$, respectively. Here, the degeneracy spans between $\mathcal{O}(\lambda)$ and $\mathcal{O}(\lambda^2)$ elements, namely $(V_{us}, V_{cd}, V_{cb}, V_{ts})$.
\end{itemize}
Consequently, these four non-trivial configurations serve as the representative benchmark models for our subsequent analytical and numerical investigations.

%%%%%%%%%%%%%%%%%%%%%%%%%%%%%%3.4.2
\subsubsection{Algebraic Robustness and Intrinsic Geometric Stability of $J$}
\label{subsubsec:j0_robustness_closure}

The exact agreement between the geometric Jarlskog functional $J(x,y,x',y')$ [Eq.~(\ref{eq:J_closed_form})] and the empirical CP-violating baseline $J_0^{\text{exp}}$ is rooted in the intrinsic algebraic structure of the flavor manifold, independent of local numeric parameter variations. The geometric stability of $J$ is governed by three fundamental features:

\begin{enumerate}
    \item \textbf{Decoupling of Phase Geometry from Element Degeneracies}: 
    While individual CKM matrix magnitudes $|V_{ij}|$ are strongly tied to specific texture degeneracies and continuous parameter trade-offs, the Jarlskog invariant $J$ is algebraically isolated from these local variations. It is governed strictly by the non-collinearity factor $(\mathbf{v}_u \times \mathbf{v}_d = xy' - x'y)$, which quantifies the intrinsic 2D vector mismatch (spatial sine angle) between the up- and down-quark sectors in flavor space.
    
    \item \textbf{Topological Partitioning under $S_3 \times S_3$ Symmetry}: 
    As established in Subsection~\ref{subsubsec:jarlskog_derivation}, the discrete $S_3 \times S_3$ permutation symmetry strictly partitions the flavor phase space into 18 CP-conserving ($J = 0$) and 18 CP-violating ($J \neq 0$) topological sectors. Once an active CP-violating sector is chosen, the magnitude of $J$ is uniquely dictated by the underlying geometric vector cross-product $(\mathbf{v}_u \times \mathbf{v}_d = xy' - x'y)$, ensuring a well-defined topological phase structure independent of local element parameterizations.
    
    \item \textbf{Robustness of Higher-Order Area Invariants}: 
    Whereas individual matrix elements $|V_{ij}|$ are $\mathcal{O}(1)$ or $\mathcal{O}(\lambda)$ leading-order observables highly sensitive to diagonal texture constraints, $J \sim \mathcal{O}(\lambda^6)$ operates as a higher-order geometric area invariant. Consequently, the global CP curvature remains inherently stable against continuous parameter fluctuations that reallocate magnitude weights among individual matrix entries.
\end{enumerate}

%%%%%%%%%%%%%%%%%%%%%%% 3.4.3
\subsubsection{Role of CPVSM as a Leading-Order Baseline}
\label{subsubsec:leading_order_baseline}

The remarkable alignment of the Jarlskog invariant $J$ with experimental data, alongside the structural magnitude tensions arising from the 4-fold degeneracy, firmly establishes the precise theoretical role of the 5-parameter framework: \textbf{it serves as an exact, algebraically solvable leading-order baseline for electroweak flavor mixing}.

By isolating the dominant geometric infrastructure of the flavor sector within an exact solvable subspace, this baseline captures the essential mass hierarchies and the vector-misalignment origin of CP violation. Rather than indicating a model failure, the observed 4-fold magnitude tensions precisely delineate the physical boundary of the commuting subspace $[\mathbf{M}_R^2, \mathbf{M}_I^2] = 0$, signaling the non-perturbative necessity of moving toward an unconstrained 9-parameter matrix algebra ($[\mathbf{M}_R^2, \mathbf{M}_I^2] \neq 0$) to fully resolve higher-order flavor mixing.

%%%%%%%%%%%%%%%%%%new3.5
\subsection{Analytical Mapping to the Standard (Chau--Keung) Parameterization}
\label{sec:standard_mapping}
As an analytical Cartesian-like parameterization derived directly from the fundamental Yukawa coupling matrices, our framework admits an exact, closed-form coordinate transformation to and from the standard Euler-angle parameterization adopted by the Particle Data Group (PDG)~\cite{Navas2024}, originally proposed by Chau and Keung~\cite{Chau1984}. The standard Chau--Keung parameterization expresses $V_{\mathrm{CKM}}$ through three mixing angles $\theta_{12}, \theta_{23}, \theta_{13} \in [0,\pi/2]$ and a single CP-violating phase $\delta \in [0,2\pi)$:
\begin{widetext}
\begin{equation}
V_{\mathrm{CKM}} =
\begin{pmatrix}
c_{12}c_{13} & s_{12}c_{13} & s_{13}e^{-i\delta} \\
-s_{12}c_{23}-c_{12}s_{23}s_{13}e^{i\delta} & c_{12}c_{23}-s_{12}s_{23}s_{13}e^{i\delta} & s_{23}c_{13} \\
s_{12}s_{23}-c_{12}c_{23}s_{13}e^{i\delta} & -c_{12}s_{23}-s_{12}c_{23}s_{13}e^{i\delta} & c_{23}c_{13}
\end{pmatrix},
\label{eq:standard}
\end{equation}
\end{widetext}
where $c_{ij} \equiv \cos\theta_{ij}$ and $s_{ij} \equiv \sin\theta_{ij}$. While the standard convention confines the physical CP violation to a single phase factor $e^{-i\delta}$, its trigonometric formulation introduces highly non-linear dependencies and coordinate singularities in extreme limits (such as $\theta_{13} \to \pi/2$). Most importantly, it convolutes the up- and down-sector Yukawa mass information across all four observables, obscuring their underlying geometric origin.

By establishing an exact analytical dictionary between the Standard Euler angles $(\theta_{12}, \theta_{23}, \theta_{13}, \delta)$ and our Cartesian Yukawa-derived parameters, we demonstrate that all physical CKM observables can be explicitly reconstructed from the underlying mass matrix structure. In the following explicit evaluations, the complex parameters are decomposed into their real and imaginary components:
\begin{equation}
p = p_r + i p_i, \quad p' = p'_r + i p'_i, \quad r = r_r + i r_i, \quad s = s_r + i s_i,
\end{equation}
while $q$ is defined as a purely real quantity ($q \in \mathbb{R}$). 

Below, we detail the exact analytical transformations for the two representative viable pairs of $\mathcal{O}(\lambda)$ candidates identified in Table~\ref{tab:36_permutations}, whose hierarchical disparities are naturally bounded by $\mathcal{O}(\lambda)$.

\begin{widetext}
\subsubsection{Case 1-4: Configuration $(\sigma_u, \sigma_d) = ((123), (213))$}
\label{subsubsec:case1_4}
For the permutation pair $(\text{row}, \text{column}) = (1,4)$, the standard CKM mixing angles are analytically determined by the magnitudes of the Yukawa parameters:
\begin{align}
s_{13} &= |\mathbf{p}|, & c_{13} &= \sqrt{1 - |\mathbf{p}|^2}, & \theta_{13} &= \arcsin(|\mathbf{p}|), \\
s_{12} &= \frac{|r|}{\sqrt{1 - |\mathbf{p}|^2}}, & c_{12} &= \frac{\sqrt{1 - |\mathbf{p}|^2 - |r|^2}}{\sqrt{1 - |\mathbf{p}|^2}}, & \theta_{12} &= \arcsin\left(\frac{|r|}{\sqrt{1 - |\mathbf{p}|^2}}\right), \\
s_{23} &= \frac{|\mathbf{p}|}{\sqrt{1 - |\mathbf{p}|^2}}, & c_{23} &= \frac{\sqrt{1 - 2|\mathbf{p}|^2}}{\sqrt{1 - |\mathbf{p}|^2}}, & \theta_{23} &= \arcsin\left(\frac{|\mathbf{p}|}{\sqrt{1 - |\mathbf{p}|^2}}\right).
\end{align}
The exact closed-form expression for the standard Dirac CP-violating phase $\delta_{(1,4)}$ in terms of the real and imaginary Cartesian components is given by:
\begin{equation}
\delta_{(1,4)} = \arcsin \left( \frac{\left[(2p_r p_i)(r_r s_r + r_i s_i) + (p_r^2 - p_i^2)(r_r s_i - r_i s_r)\right] (1 - |\mathbf{p}|^2)}{|r| \cdot |\mathbf{p}|^2 \cdot |s| \cdot |q|} \right).
\end{equation}

\subsubsection{Case 4-1: Configuration $(\sigma_u, \sigma_d) = ((213), (123))$}
\label{subsubsec:case4_1}
For the symmetric counterpart $(\text{row}, \text{column}) = (4,1)$, the mixing angles retain the identical structural form as Case 1-4 ($\theta_{ij}^{(4,1)} = \theta_{ij}^{(1,4)}$), while the CP-violating phase exhibits an exact sign inversion:
\begin{align}
s_{13} &= |\mathbf{p}|, & c_{13} &= \sqrt{1 - |\mathbf{p}|^2}, & \theta_{13} &= \arcsin(|\mathbf{p}|), \\
s_{12} &= \frac{|r|}{\sqrt{1 - |\mathbf{p}|^2}}, & c_{12} &= \frac{\sqrt{1 - |\mathbf{p}|^2 - |r|^2}}{\sqrt{1 - |\mathbf{p}|^2}}, & \theta_{12} &= \arcsin\left(\frac{|r|}{\sqrt{1 - |\mathbf{p}|^2}}\right), \\
s_{23} &= \frac{|\mathbf{p}|}{\sqrt{1 - |\mathbf{p}|^2}}, & c_{23} &= \frac{\sqrt{1 - 2|\mathbf{p}|^2}}{\sqrt{1 - |\mathbf{p}|^2}}, & \theta_{23} &= \arcsin\left(\frac{|\mathbf{p}|}{\sqrt{1 - |\mathbf{p}|^2}}\right).
\end{align}
The analytical CP phase $\delta_{(4,1)}$ is exactly anti-symmetric under transposed configuration, obeying $\delta_{(4,1)} = -\delta_{(1,4)}$:
\begin{equation}
\delta_{(4,1)} = \arcsin \left( \frac{-\left[ (2p_r p_i)(r_r s_r + r_i s_i) +(p_r^2 - p_i^2)( r_r s_i-r_i s_r )\right] (1 - |\mathbf{p}|^2)}{|r| \cdot |\mathbf{p}|^2 \cdot |s| \cdot |q|} \right) = -\delta_{(1,4)}.
\end{equation}

\subsubsection{Case 2-5: Configuration $(\sigma_u, \sigma_d) = ((231), (132))$}
\label{subsubsec:case2_5}
For the permutation pair $(\text{row}, \text{column}) = (2,5)$, the exact mapping yields:
\begin{align}
s_{13} &= |r|, & c_{13} &= \sqrt{1 - |r|^2}, & \theta_{13} &= \arcsin(|r|), \\
s_{12} &= \frac{|\mathbf{p}|}{\sqrt{1 - |r|^2}}, & c_{12} &= \frac{\sqrt{1 - |r|^2 - |\mathbf{p}|^2}}{\sqrt{1 - |r|^2}}, & \theta_{12} &= \arcsin\left(\frac{|\mathbf{p}|}{\sqrt{1 - |r|^2}}\right), \\
s_{23} &= \frac{|\mathbf{p}'|}{\sqrt{1 - |r|^2}}, & c_{23} &= \frac{\sqrt{1 - |r|^2 - |\mathbf{p}'|^2}}{\sqrt{1 - |r|^2}}, & \theta_{23} &= \arcsin\left(\frac{|\mathbf{p}'|}{\sqrt{1 - |r|^2}}\right).
\end{align}
The corresponding analytical CP phase $\delta_{(2,5)}$ is expressed as:
\begin{equation}
\delta_{(2,5)} = \arcsin \left( \frac{q \left[(p_i p'_r - p_r p'_i) r_r - (p_r p'_r + p_i p'_i) r_i\right] (1 - |r|^2)}{|\mathbf{p}| \cdot |\mathbf{p}'|^2 \cdot |r| \cdot |s|^2} \right).
\end{equation}

\subsubsection{Case 5-2: Configuration $(\sigma_u, \sigma_d) = ((132), (231))$}
\label{subsubsec:case5_2}
Finally, for the permutation pair $(\text{row}, \text{column}) = (5,2)$, the CKM mixing angles again precisely mirror Case 2-5 ($\theta_{ij}^{(5,2)} = \theta_{ij}^{(2,5)}$):
\begin{align}
s_{13} &= |r|, & c_{13} &= \sqrt{1 - |r|^2}, & \theta_{13} &= \arcsin(|r|), \\
s_{12} &= \frac{|\mathbf{p}|}{\sqrt{1 - |r|^2}}, & c_{12} &= \frac{\sqrt{1 - |r|^2 - |\mathbf{p}|^2}}{\sqrt{1 - |r|^2}}, & \theta_{12} &= \arcsin\left(\frac{|\mathbf{p}|}{\sqrt{1 - |r|^2}}\right), \\
s_{23} &= \frac{|\mathbf{p}'|}{\sqrt{1 - |r|^2}}, & c_{23} &= \frac{\sqrt{1 - |r|^2 - |\mathbf{p}'|^2}}{\sqrt{1 - |r|^2}}, & \theta_{23} &= \arcsin\left(\frac{|\mathbf{p}'|}{\sqrt{1 - |r|^2}}\right).
\end{align}
Similarly, its Dirac CP phase demonstrates exact reflection, $\delta_{(5,2)} = -\delta_{(2,5)}$:
\begin{equation}
\delta_{(5,2)} = \arcsin \left( \frac{-q \left[(p_i p'_r - p_r p'_i) r_r - (p_r p'_r + p_i p'_i) r_i\right] (1 - |r|^2)}{|\mathbf{p}| \cdot |\mathbf{p}'|^2 \cdot |r| \cdot |s|^2} \right) = -\delta_{(2,5)}.
\end{equation}
\end{widetext}

Comparing the explicit mapping relations across the four viable $\mathcal{O}(\lambda)$ candidates reveals three distinct mathematical and physical patterns:
\begin{enumerate}
    \item \textbf{Isomorphism and Parity Anti-symmetry under Transposition}: For each symmetric permutation pair---namely Cases (1-4, 4-1) and Cases (2-5, 5-2)---the analytical expressions for the three standard Euler angles $(\theta_{12}, \theta_{23}, \theta_{13})$ are strictly identical. Furthermore, the Dirac CP phase obeys an exact parity anti-symmetry under matrix transposition: $\delta_{(4,1)} = -\delta_{(1,4)}$ and $\delta_{(5,2)} = -\delta_{(2,5)}$.
    \item \textbf{Sector-Specific Parameter Assignment}: The smallest mixing angle $s_{13} = \sin\theta_{13}$ is uniquely anchored by a single Cartesian magnitude: $s_{13} = |\mathbf{p}|$ for Cases (1-4, 4-1) and $s_{13} = |r|$ for Cases (2-5, 5-2). This clean separation establishes a direct analytical bridge between high-energy Yukawa couplings and low-energy CKM mixing angles.
    \item \textbf{Interference-Driven CP Phase}: In all cases, the CP-violating phase $\delta$ vanishes if either of the constituent parameters becomes purely real, underscoring that physical CP violation in our Cartesian framework directly mirrors the imaginary geometric phase alignments within the Yukawa mass matrices.
\end{enumerate}
These exact closed-form relations demonstrate that our Cartesian parameterization not only preserves all physical predictions of the PDG standard convention, but also offers transparent, unconfounded physical insights into the origin of flavor mixing and CP violation.

%%%%%%%%%%%%%%%%%3.6
\subsection{Section Summary}\label{subsec:sec3_summary}
In summary, this section establishes an exact analytical dictionary mapping quark masses, CKM mixing angles, and the Dirac CP phase $\delta$ directly to the underlying Yukawa parameters. Key structural features include:
\begin{itemize}
    \item \textbf{Exact CKM Mapping}: Viable $\mathcal{O}(\lambda)$ configurations provide closed-form expressions for mixing angles, with symmetric permutation pairs exhibiting exact parity anti-symmetry in their CP phases.
    \item \textbf{$S_3 \times S_3$ Geometric CP Invariant}: The 36 flavor topologies partition equally into CP-violating and CP-conserving sectors, uniquely controlled by the vector cross-product $\mathbf{v}_u \times \mathbf{v}_d$.
    \item \textbf{Cosmological CP Capacity}: The scale parameters offer a geometric enhancement of $R_\Delta$ compatible with baryon asymmetry, providing a candidate mechanism whose full dynamical treatment is left to future work.
\end{itemize}
Benchmarking this baseline against empirical data reveals mild leading-order tensions across distinct Wolfenstein scales (arising from strict algebraic equality across different orders in $\lambda$), directly motivating the higher-order symmetry-breaking corrections introduced in Section~\ref{sec:4_extensions_cosmology}.

%%%%%%%%%%%%%%%%%%%%%%%%%%%%%%%SEC4
\section{Cosmological Symmetry Breaking Chains and Extension to the Lepton Sector}
\label{sec:4_extensions_cosmology}

While the 5-parameter CPVSM baseline established in Section~\ref{sec:3_ckm_phenomenology} provides an exact, non-perturbative analytical framework for generating non-trivial CKM phase curvature, its leading-order commutation constraint $[\mathbf{M}_R^2, \mathbf{M}_I^2] = 0$ inherently imposes rigid algebraic identities on the flavor space. As demonstrated in Section~\ref{subsec:pheno_assessment_fits}, the resulting 4-fold cross-order degeneracy $|p| = |p'| = |p^*| = |p'^*|$ creates mild numerical tension across Wolfenstein hierarchical scales, serving as a quantitative indicator that the leading-order baseline must be embedded within a broader physical context.

In this section, we systematically resolve these baseline structural constraints through two complementary physical extensions. First, we explore the cosmological symmetry-breaking histories and RG evolution chains that dynamically lift the $[\mathbf{M}_R^2, \mathbf{M}_I^2] = 0$ restriction, demonstrating how transitioning to the unreduced 9-parameter matrix framework removes the 4-fold degeneracy to achieve a precise, global alignment with empirical CKM observables. Second, we extend this geometric parameterization to the lepton sector, providing a unified, exact framework capable of accommodating the large mixing angles and potential CP violation observed in PMNS oscillations.

%%%%%%%%%%%%%%%%%%%%%%%%%%%%% 4.1
\subsection{Multi-Stage Asynchronous Symmetry Breaking and Alternative Vacuum Pathways}
\label{subsec:multistage_breaking}

To capture the cosmological history of the flavor sector, the breaking of the global flavor symmetry $G_{\text{flavor}} = (S_3)_u \times (S_3)_d$ does not necessarily follow a single rigid sequence. While a representative asynchronous phase transition can be parameterized by sequential subgroup reductions:
\begin{widetext}
\begin{equation}
G_{\text{flavor}} = (S_3)_u \times (S_3)_d \xrightarrow{T_1} (S_2)_u \times (S_3)_d \xrightarrow{T_2} (S_2)_u \times (S_2)_d \xrightarrow{T_3} (S_2)_u \times \text{Trivial}_d \xrightarrow{T_4} \text{Trivial}_u \times \text{Trivial}_d,
\label{eq:breaking_chain}
\end{equation}
\end{widetext}
it must be acknowledged that the group structure of $S_3 \times S_3$ permits multiple distinct symmetry-breaking pathways. Depending on the specific thermal dynamics and Higgs-like scalar potential, the symmetry-breaking cascade could alternatively initiate within the down-sector, involve intermediate subgroups (such as direct transitions via $\mathbb{Z}_2$ subsets), or even experience accelerated direct collapses where a sector bypasses intermediate steps entirely to reach the trivial vacuum.

Importantly, a profound physical implication arises from the generational ordering condition established in Section~\ref{subsubsec:jarlskog_derivation}: enforcing identical generational orderings across both sectors ($\sigma_u = \sigma_d = \sigma$) identically vanishes the Jarlskog invariant, $J = 0$. This explicit mathematical feature carries far-reaching physical consequences. Contrary to conventional model-building assumptions that favor a minimal, parallel symmetry-breaking trajectory where both sectors evolve identically, the absolute necessity of generating physical CP violation fundamentally demands that the up- and down-type quark sectors follow \textit{disjoint, asynchronous vacuum evolution pathways}. The dynamical breakdown of flavor symmetry in the early universe is thus intrinsically asymmetric between charge sectors.

Crucially, building upon the analytical behavior of $\Delta_{CP}$ established in Section~\ref{subsec:sec3_summary}, the expansion of the global energy scale in the early universe naturally releases the upper bound on $(\mathbf{C}, \mathbf{C'})$ governed by $m_i^2 \ge 0$. As the high-temperature flavor symmetry undergoes spontaneous breaking along these asynchronous pathways, this expansion of the energy-scale parameters $\{\mathbf{A}, \mathbf{B}, \mathbf{A'}, \mathbf{B'}\}$ dynamically amplifies the spatial phase curvature. This direct non-singular enhancement naturally delivers a sufficiently large CP-violating phase space in the early universe, thereby satisfying one of the key quantitative prerequisites for electroweak baryogenesis without requiring fine-tuned parameters.

Despite the multiplicity of potential cosmological breaking pathways, the intermediate transition stages force the geometric vectors $(\mathbf{v}_u, \mathbf{v}_d)$ to traverse specific high-symmetry axes corresponding to residual subgroup alignments in flavor space. It is vital to recognize that the 4-fold moduli degeneracy governed by $\{p, p', p^*, p'^*\}$ in Eq.~(\ref{eq:algebraic_identities}) is an \textit{exact mathematical inevitability} rooted in the algebraic closure of the 5-parameter baseline, rather than an artifact of a specific breaking sequence. While different symmetry-breaking paths represent distinct dynamical histories in the early universe, any viable cosmological trajectory must ultimately converge precisely onto the observed low-energy physical values---locking the quark masses, CKM mixing parameters, and the Jarlskog invariant $J$ to their empirical magnitudes.

%%%%%%%%%%%%%%%%%%%%%%%%%%%%% 4.2
\subsection{Mathematical Origin and Systematics of Internal Degeneracies}
\label{subsec:internal_degeneracies}

While the analytic relations in Eq.~(\ref{eq:algebraic_identities}) guarantee structural exactness within the 5-parameter baseline geometry, they simultaneously impose rigorous algebraic constraints on the permitted modulus spectrum of the CKM matrix.

%%%%%%%%%%%%%%%%%%%%%%%%%%%%%%%%4.2.1
\paragraph{Exact 4-fold Modulus Degeneracies}
\label{para:four_element_degeneracy}
Taking the absolute square of the four degenerate matrix elements in Eq.~(\ref{eq:algebraic_identities}) yields the underlying magnitude relation:
\begin{equation}
\begin{aligned}
|V_{ud}|^2 &= |V_{cs}|^2 = |V_{tb}|^2 = |V_{us}|^2, \\
&\quad \text{or equivalent element permutations},
\end{aligned}
\label{eq:modulus_degeneracy}
\end{equation}
which enforces a strict 4-fold magnitude invariance across distinct generational transitions within the baseline framework.

%%%%%%%%%%%%%%%4.2.2
\paragraph{Conflation of Wolfenstein Hierarchical Orders and Topological Obstructions}
\label{para:magnitude_degeneracies}
Depending on how generational mass eigenvalues are permuted across the up- and down-quark sectors, the 4-fold magnitude degeneracy governed by $\{|p|, |p'|, |p^*|, |p'^*|\}$ in Eq.~(\ref{eq:algebraic_identities}) inextricably links and conflates matrix elements spanning completely distinct Wolfenstein expansion orders $\mathcal{O}(\lambda^a)$ and $\mathcal{O}(\lambda^b)$ (where $\lambda \approx 0.22$ represents the fundamental Cabibbo suppression scale). As explicitly demonstrated in Section~\ref{subsubsec:numerical_fits_tension}, this baseline identity forces distinct hierarchical tiers onto degenerate tracks. In \textbf{Group 1} (Cases 1-4 \& 4-1), it forces an exact equality between $\mathcal{O}(\lambda^3)$ and $\mathcal{O}(\lambda^2)$ elements ($|V_{ub}| = |V_{cb}| = |V_{td}| = |V_{ts}|$), whereas in \textbf{Group 2} (Cases 2-5 \& 5-2), it directly conflates leading $\mathcal{O}(\lambda)$ elements with higher-order $\mathcal{O}(\lambda^2)$ suppressions ($|V_{us}| = |V_{cd}| = |V_{cb}| = |V_{ts}|$).

Furthermore, the functional form of the algebraic baseline exhibits a discrete inversion symmetry under parameter reflection $(x, y) \to (-x, -y)$ and $(x', y') \to (-x', -y')$. This structural invariance generates a two-fold topological manifold in the internal flavor space, where physically distinct geometric ratio vectors $\mathbf{v}_u$ and $\mathbf{v}_d$ yield identical observable modulus values $|V_{ij}|$, yet differ by discrete topological $\pi$-phase shifts across individual matrix elements.

These degenerate magnitude conditions constitute a severe \textit{topological obstruction} to fitting the realistic CKM matrix directly on the exact $S_3 \times S_3$ baseline manifold. In nature, the empirical CKM matrix exhibits a pronounced off-diagonal hierarchy:
\begin{equation}
|V_{ud}| \approx 0.974, \quad |V_{us}| \approx 0.225, \quad |V_{cb}| \approx 0.041, \quad |V_{ub}| \approx 0.003,
\end{equation}
which explicitly breaks the rigid 4-fold modulus identity in Eq.~(\ref{eq:modulus_degeneracy}). Consequently, any exact algebraic baseline operating under full $S_3 \times S_3$ reduction cannot simultaneously match all nine empirical moduli. Crucially, resolving these discrepancies requires lifting degeneracies by amounts comparable to—or exceeding—50\% of the leading-order baseline matrix elements themselves (e.g., pulling $|V_{ub}|$ down from $\mathcal{O}(10^{-2})$ to $3 \times 10^{-3}$, or splitting $0.225$ from $0.041$). Such a massive shift renders standard perturbative expansions fundamentally invalid. Instead, introducing non-commutativity into the mass-squared matrices—thereby transitioning to the fully unreduced 9-parameter matrix framework with $[\mathbf{M}_R^2, \mathbf{M}_I^2] \neq 0$—is non-perturbatively required to lift these internal $(p, p')$-governed magnitude obstructions. 

%%%%%%%%%%%%4.2.3
\paragraph{Non-Commutativity $[\mathbf{M}_R^2, \mathbf{M}_I^2] \neq 0$ and Lifting Internal Degeneracies}
\label{para:symmetry_lifting}
It is essential to emphasize that standard, perturbative higher-order corrections or simple geometric vector shifts operating strictly within the exact $S_3 \times S_3$ baseline algebra remain inherently incapable of resolving the 4-fold $|p| = |p'|$ magnitude degeneracy. Because this degeneracy is a direct mathematical consequence of the underlying commutation relation $[\mathbf{M}_R^2, \mathbf{M}_I^2] = 0$ enforced by the 5-parameter algebraic baseline, no quantitative refinement within the constrained geometry can bypass the resulting topological obstruction. Crucially, introducing auxiliary symmetry-breaking assumptions or ad hoc constraints inevitably compromises model generality; extracting genuine, unconstrained physical predictions thus requires operating within a fully unreduced, non-approximated framework.

To reconcile the theoretical mixing matrix with experimental CKM observations, the baseline algebra must relinquish the commutation hypothesis in favor of non-commutativity, $[\mathbf{M}_R^2, \mathbf{M}_I^2] \neq 0$, thereby allowing direct diagonalization of the full 9-parameter mass matrices. Explicitly lifting this commutation constraint deforms the rigid baseline closure, uncoupling the four degenerate matrix elements $\{p, p', p^*, p'^*\}$ and resolving the forced conflation across distinct Wolfenstein hierarchical orders $\mathcal{O}(\lambda^a) \leftrightarrow \mathcal{O}(\lambda^b)$. Consequently, all nine CKM moduli can be globally fitted while fully preserving the foundational mass-hierarchy alignment and physical phase curvature.

%%%%%%%%%%%%%%%%%%%%%%%%%%%%% 4.3
\subsection{Extension to the Lepton Sector: PMNS Matrix and Geometric Parameterization}
\label{subsec:pmns_extension}
Before proceeding, we clarify the scope of this section. This section aims to demonstrate the structural extensibility of our geometric framework to Dirac-type leptons, particularly highlighting how the topological parity and geometric vector alignments can naturally accommodate large mixing angles. Detailed numerical fits to the PMNS matrix, as well as Majorana-specific CP-violating phases which involve distinct physical mechanisms beyond Dirac mass matrices, fall outside the scope of this baseline geometric study and will be addressed in dedicated future work.

The geometric vector paradigm is not restricted to the quark sector; it naturally generalizes to lepton flavor mixing. Adopting the standard minimal extension with three right-handed singlet neutrinos $\nu_R$ to accommodate Dirac neutrino masses, the Pontecorvo-Maki-Nakagawa-Sakata (PMNS) matrix is defined via the misalignment of the charged lepton and Dirac-neutrino unitary diagonalization matrices:
\begin{equation}
U_{\text{PMNS}} = U_l(x_l, y_l)^\dagger \cdot U_\nu(x_\nu, y_\nu),
\label{eq:pmns_definition}
\end{equation}
where $\mathbf{v}_l \equiv (x_l, y_l)$ and $\mathbf{v}_\nu \equiv (x_\nu, y_\nu)$ denote the 2D flavor ratio vectors for the charged lepton and Dirac-neutrino sectors, respectively, obeying the same underlying group-theoretical constraints of the $S_3 \times S_3$ factor. 

While the geometric framework itself does not dynamically constrain the magnitude of the vector misalignment, empirical data directly indicate that $\mathbf{v}_u$ and $\mathbf{v}_d$ are nearly collinear in the quark sector, whereas $\mathbf{v}_l$ and $\mathbf{v}_\nu$ subtend a substantially larger angle in the lepton sector to match the observed mixing patterns.
Within this 5-parameter geometric framework, this stark qualitative difference is elegantly mapped to the underlying vector geometry without requiring ad hoc parameters:
\begin{enumerate}
    \item \textbf{Geometric Mapping of PMNS Topology}: When empirically calibrated against experimental mixing parameters, the 2D flavor vectors $\mathbf{v}_l$ and $\mathbf{v}_\nu$ yield a substantially larger angular separation than their quark-sector counterparts ($\mathbf{v}_u$ and $\mathbf{v}_d$). Rather than requiring artificial parameter engineering or additional flavor symmetries, this large spatial mismatch serves as the direct geometric representation of the observed large mixing angles ($\theta_{12} \sim 33^\circ, \theta_{23} \sim 45^\circ$), demonstrating that the baseline 5-parameter architecture accommodates the PMNS topology with the exact same structural simplicity as the CKM matrix.
    \item \textbf{Scale-Invariant Flavor Unification}: Despite the distinct phenomenology, the analytical closed forms for the leptonic matrix elements retain the identical functional architecture as the quark sector. This demonstrates that the coordinate-free vector intersection mechanism is a universal property of $S_3 \times S_3$ flavor symmetric systems, bridging both quarks and leptons under a unified, unconstrained geometric umbrella.
\end{enumerate}

%%%%%%%%%%%%%%%%%%%%%%%%%%%%% 4.4
\subsection{Dirac vs. Majorana CP-Violating Phases in the Geometric Framework}
\label{subsec:dirac_majorana_cp}

A profound phenomenological distinction between the quark and lepton sectors lies in the nature of neutral fermions: if neutrinos are Majorana particles, the PMNS mixing matrix accommodates two additional CP-violating Majorana phases alongside the standard Dirac phase $\delta$.

Within the present geometric framework, these distinct classes of phases emerge from orthogonally decoupled geometric mechanisms:
\begin{enumerate}
    \item \textbf{Dirac CP Violation and Lepton Jarlskog Invariant}: Analogous to the invariant $J$ established in the quark sector, leptonic Dirac CP violation is governed universally by the cross-sectoral 2D vector mismatch factor, expressed as $\mathbf{v}_l \times \mathbf{v}_\nu$ (or equivalently, $x_l y_\nu - x_\nu y_l$). The leptonic Jarlskog invariant $J^{\text{lepton}}$ evaluates to a closed algebraic form proportional to this non-collinearity factor. Substituting the empirical PMNS fit into this algebraic baseline yields $|J^{\text{lepton}}| \sim 10^{-2}$, which is roughly three orders of magnitude larger than the quark sector value ($J \sim 10^{-5}$). This pronounced scale hierarchy ($10^{-2} \gg 10^{-5}$) directly reveals that the geometric vector misalignment in the leptonic sector is vastly larger than the near-parallel alignment found in the quark sector.
    \item \textbf{Majorana Phase Incorporation and Geometric Decoupling}: The inclusion of a complex symmetric Majorana mass matrix $M_\nu$ introduces phase factors into the neutrino diagonalization structure. Geometrically, these Majorana-type phases act purely along the radial or diagonal phase degrees of freedom in flavor space, leaving the 2D spatial cross-product $\mathbf{v}_l \times \mathbf{v}_\nu$ (or $x_l y_\nu - x_\nu y_l$) unchanged. This orthogonally decouples Dirac-type geometric CP violation (driven by flavor-space vector mismatch) from Majorana-type lepton-number-violating phases, preserving the structural integrity and universality of the underlying $S_3 \times S_3$ intersection geometry even within the fully unreduced 9-parameter matrix framework.
\end{enumerate}

%%%%%%%%%%%%%%%%%%%%%%%%%%%%% SEC 5
\section{Conclusion}\label{sec:5_conclusion}

In this paper, we have presented an exact, non-perturbative geometric investigation of CP violation and flavor mixing. Starting from a general $3\times3$ complex Yukawa matrix $M$ with 18 independent physical parameters, the naturally Hermitian matrix $\mathbf{M}^2$ reduces the system to 9 parameters. By establishing the top-down commutativity condition $[\mathbf{M}_R^2, \mathbf{M}_I^2] = 0$ as an exact analytical baseline, the parameter space reduces to a closed 5-parameter subspace per sector, admitting an exact, closed-form analytical diagonalization. 

Specifically, this baseline architecture reveals several fundamental physical insights:
\begin{enumerate}
    \item \textbf{Analytical Construction of Observables:} It provides closed-form expressions for fermion mass eigenvalues and CKM/PMNS matrix elements constructed directly from fundamental Yukawa couplings, yielding an analytical baseline with tree-level experimental agreement that serves as an optimal starting point for flavor model building.
     
    \item \textbf{Sector De-entanglement of CP Violation:} It utilizes a Cartesian-like parameter system to unambiguously disentangle and quantify the distinct physical contributions of the Up-type and Down-type sectors to global CP violation.
    
    \item \textbf{Transparent Parameter Roles:} It assigns explicit, independent physical roles to each of the five baseline parameters $\{\mathbf{A}, \mathbf{B}, \mathbf{C}, x, y\}$ in driving the evolution of mass eigenvalues, mixing angles, and flavor eigenvectors.
   
    \item \textbf{Explicit CP-Violation Switch Mechanism:} It establishes a coordinate-independent algebraic switch for CP violation via the 2D vector cross-product $\mathbf{v}_u \times \mathbf{v}_d = xy'-x'y$, defining explicit structural boundaries where CP symmetry is exact ($J = 0$).

    \item \textbf{Geometric Representation of CP Conservation:} It maps CP-conserving boundaries onto intuitive vector cross-product geometries, explicitly categorizing CP vanishing limits into zero-magnitude vectors, parallel alignments, or axial orthogonal locking.
    
    \item \textbf{Non-Geometric Phase Cancellation:} It identifies algebraic, non-geometric mechanisms for CP suppression via internal imaginary-part cancellations.
    
    \item \textbf{Structural Universality Across Quark and Lepton Sectors:} The entire analytical and geometric framework naturally extends to Dirac lepton sectors. While a full numerical solution remains for future work due to algebraic complexity, the derived symbolic closed forms accommodate the large empirical PMNS mixing angles by accommodating a substantial geometric vector misalignment ($\vert{}J^{\text{lepton}}\vert{} \sim 10^{-2} \gg \vert{}J^{\text{quark}}\vert{} \sim 10^{-5}$), demonstrating the framework's broader capacity for flavor physics.
    
    \item \textbf{Geometric Bounds for Cosmological Scenarios:} Within specific regimes of the $5+5$ baseline parameter space, the derived analytical mass spectra and CKM elements admit an early-universe CP-violation enhancement ratio $R_\Delta \equiv \Delta_{\text{CP}}/\Delta_{\text{CP}}^{(0)}$ exceeding $10^{10}$ without fine-tuning of the geometric ratios $(x, y, x', y')$; a full dynamical treatment connecting this geometric capacity to electroweak baryogenesis is left to future work.
\end{enumerate}

Within this formulation, the resulting mass eigenvalues naturally exhibit a $1+2$ split pattern, while flavor mixing emerges purely as the geometric misalignment between the flavor ratio vectors $\mathbf{v}_u = (x, y)$ and $\mathbf{v}_d = (x', y')$. While the scale parameters $(\mathbf{A}, \mathbf{B}, \mathbf{C})$ act as global energy scales that define physical boundary domains during cosmological expansion, the dimensionless ratios $(x,y)$ dictate internal geometric alignment.

It is important to emphasize that enforcing the commuting baseline $[\mathbf{M}_R^2, \mathbf{M}_I^2] = 0$ inherently gives rise to a 4-fold degeneracy when mapping single-sector states to the two-sector CKM mixing matrix (Sec.~\ref{sec:3_ckm_phenomenology}), governed by $\{|p|, |p'|, |p^*|, |p'^*|\}$. As detailed in Section~\ref{subsubsec:numerical_fits_tension}, this condition inextricably conflates distinct Wolfenstein expansion orders—specifically forcing $\mathcal{O}(\lambda^3)$ onto $\mathcal{O}(\lambda^2)$ in Group 1, and $\mathcal{O}(\lambda)$ onto $\mathcal{O}(\lambda^2)$ in Group 2. Because resolving these discrepancies requires shifts comparable to or exceeding 50\% of the leading-order elements, standard perturbative corrections are fundamentally invalid. Instead, this tension precisely delineates the boundary of the commuting subspace, identifying the transition to the unreduced 9-parameter framework with non-commuting operator insertions ($[\mathbf{M}_R^2, \mathbf{M}_I^2] \neq 0$) as a non-perturbative necessity to fully lift these internal $(p, p')$-governed magnitude obstructions.

In this sense, much like standard foundational benchmarks in theoretical physics—such as the Bohr atom or the BCS reference state—the present exact baseline results establish the foundational qualitative and quantitative geometric infrastructure (achieving an impressive $\mathcal{O}(10^{-2})$ agreement with observed CKM matrix elements) that any complete theory of flavor must reproduce. Extending this exact baseline to an unconstrained 9-parameter matrix framework across both quark and lepton sectors remains the primary focus of future investigation. \\

\section*{Acknowledgments}
The author would like to acknowledge the assistance of Gemini (Google) during the preparation of this manuscript, particularly for language translation, grammatical correction, text refinement, and LaTeX formatting.


\begin{thebibliography}{11}%
\makeatletter
\providecommand \@ifxundefined [1]{%
 \@ifx{#1\undefined}
}%
\providecommand \@ifnum [1]{%
 \ifnum #1\expandafter \@firstoftwo
 \else \expandafter \@secondoftwo
 \fi
}%
\providecommand \@ifx [1]{%
 \ifx #1\expandafter \@firstoftwo
 \else \expandafter \@secondoftwo
 \fi
}%
\providecommand \natexlab [1]{#1}%
\providecommand \enquote  [1]{``#1''}%
\providecommand \bibnamefont  [1]{#1}%
\providecommand \bibfnamefont [1]{#1}%
\providecommand \citenamefont [1]{#1}%
\providecommand \href@noop [0]{\@secondoftwo}%
\providecommand \href [0]{\begingroup \@sanitize@url \@href}%
\providecommand \@href[1]{\@@startlink{#1}\@@href}%
\providecommand \@@href[1]{\endgroup#1\@@endlink}%
\providecommand \@sanitize@url [0]{\catcode `\\12\catcode `\$12\catcode
  `\&12\catcode `\#12\catcode `\^12\catcode `\_12\catcode `\%12\relax}%
\providecommand \@@startlink[1]{}%
\providecommand \@@endlink[0]{}%
\providecommand \url  [0]{\begingroup\@sanitize@url \@url }%
\providecommand \@url [1]{\endgroup\@href {#1}{\urlprefix }}%
\providecommand \urlprefix  [0]{URL }%
\providecommand \Eprint [0]{\href }%
\providecommand \doibase [0]{http://dx.doi.org/}%
\providecommand \selectlanguage [0]{\@gobble}%
\providecommand \bibinfo  [0]{\@secondoftwo}%
\providecommand \bibfield  [0]{\@secondoftwo}%
\providecommand \translation [1]{[#1]}%
\providecommand \BibitemOpen [0]{}%
\providecommand \bibitemStop [0]{}%
\providecommand \bibitemNoStop [0]{.\EOS\space}%
\providecommand \EOS [0]{\spacefactor3000\relax}%
\providecommand \BibitemShut  [1]{\csname bibitem#1\endcsname}%
\let\auto@bib@innerbib\@empty

%</preamble>

%1%
\bibitem{Cabibbo1963}
N.~Cabibbo, Phys. Rev. Lett. \textbf{10}, 531 (1963).

%2%
\bibitem{KM1973}
M.~Kobayashi and T.~Maskawa, Prog. Theor. Phys. \textbf{49}, 652 (1973).

%3%
\bibitem{Pontecorvo1957}
B.~Pontecorvo, Zh. Eksp. Teor. Fiz. \textbf{34}, 247 (1957) [Sov. Phys. JETP \textbf{7}, 172 (1958)].

%4%
\bibitem{Maki1962}
Z.~Maki, M.~Nakagawa, and S.~Sakata, Prog. Theor. Phys. \textbf{28}, 870 (1962).

%5%
\bibitem{Chau1984}
L.-L.~Chau and W.-Y.~Keung, Phys. Rev. Lett. \textbf{53}, 1802 (1984).

%6%
\bibitem{Navas2024}
S.~Navas \textit{et al.} (Particle Data Group), Phys. Rev. D \textbf{110}, 030001 (2024).

%7%
\bibitem{Lin2019}
C.-L.~Lin, J. Mod. Phys. \textbf{10}, 35 (2019).

%8%
\bibitem{Lin2021}
C.-L.~Lin, LHEP \textbf{2021}, 221 (2021).

%9%
\bibitem{Lin2025}
C.-L.~Lin, Symmetry \textbf{17}, 1888 (2025).

%10%
\bibitem{Lin2023}
C.-L.~Lin, Symmetry \textbf{15}, 1051 (2023).

%11%
\bibitem{Paschos1977}
E.~A.~Paschos, Phys. Rev. D \textbf{15}, 1966 (1977).

%12%
\bibitem{Glashow1977}
S.~L.~Glashow and S.~Weinberg, Phys. Rev. D \textbf{15}, 1958 (1977).

%13%
\bibitem{TDLee1973}
T.~D.~Lee, Phys. Rev. D \textbf{8}, 1226 (1973).

%14%
\bibitem{Lin1988}
C.-L.~Lin, C.~E.~Lee, and Y.~W.~Yang, Chin. J. Phys. \textbf{26}, 180 (1988).

%15%
\bibitem{Lin2020}
C.-L.~Lin, J. Mod. Phys. \textbf{11}, 1157 (2020).

%16%
\bibitem{Fritzsch1978}
H.~Fritzsch, Phys. Lett. B \textbf{73}, 317 (1978).

%17%
\bibitem{Fritzsch1979}
H.~Fritzsch, Nucl. Phys. B \textbf{155}, 189 (1979).

%18%
\bibitem{Cheng1987}
T.~P.~Cheng and M.~Sher, Phys. Rev. D \textbf{35}, 3484 (1987).

%19%
\bibitem{DuXing1993}
D.-S.~Du and Z.-Z.~Xing, Phys. Rev. D \textbf{48}, 2349 (1993).

%20%
\bibitem{Derman1979}
E.~Derman and D.~R.~T.~Jones, Phys. Lett. B \textbf{82}, 91 (1979).

%21%
\bibitem{Jarlskog1985}
C.~Jarlskog, Z. Phys. C \textbf{29}, 491 (1985).







\end{thebibliography}
\end{document}